\documentclass[journal,twoside,web]{ieeecolor}
\let\labelindent\relax
\usepackage{generic}

\usepackage{cite}
\usepackage{amsmath,amssymb,amsfonts}
\usepackage{algorithmic}
\usepackage{graphicx}
\usepackage{textcomp}
\usepackage{xcolor}
\usepackage{mathtools}
\usepackage{xcolor}
\usepackage[utf8]{inputenc}
\usepackage[T1]{fontenc}
\usepackage{url}
\usepackage{subfigure}
\usepackage{enumitem}
\usepackage{threeparttable}
\usepackage{multirow}
\usepackage{comment}
\usepackage{epstopdf}
\usepackage{multirow}
\usepackage{booktabs}

\begin{document}


\title{Channel Modeling Framework for Both Communications and Bistatic Sensing Under 3GPP Standard}

\author{Chenhao~Luo,~Aimin~Tang,~\IEEEmembership{Member,~IEEE,}~Fei~Gao,~\IEEEmembership{Senior~Member,~IEEE,}~Jianguo~Liu,~\IEEEmembership{Member,~IEEE,}~and~Xudong~Wang,~\IEEEmembership{Fellow,~IEEE}
\thanks{The Chenhao Luo, Aimin Tang, and Xudong Wang are with the University of Michigan-Shanghai Jiao Tong University Joint Institute, Shanghai Jiao Tong University. Fei Gao and Jianguo Liu are with the Noika Bell Labs China. Corresponding author: Aimin Tang (email: tangaiming@sjtu.edu.cn).
Part of this paper was presented at the 2024 IEEE Vehicular Technology Conference (VTC2024-Spring), Singapore, June 24-27, 2024 \cite{luo2024vtc}.}
}


\maketitle

\begin{abstract}
Integrated sensing and communications (ISAC) is considered a promising technology in the B5G/6G networks. The channel model is essential for an ISAC system to evaluate the communication and sensing performance. Most existing channel modeling studies focus on the monostatic ISAC channel. In this paper, the channel modeling framework for bistatic ISAC is considered. The proposed channel modeling framework extends the current 3GPP channel modeling framework and ensures the compatibility with the communication channel model. To support the bistatic sensing function, several key features for sensing are added. First, more clusters with weaker power are generated and retained to characterize the potential sensing targets. Second, the target model can be either deterministic or statistical, based on different sensing scenarios. Furthermore, for the statistical case, different reflection models are employed in the generation of rays, taking into account spatial coherence. The effectiveness of the proposed bistatic ISAC channel model framework is validated by both ray tracing simulations and experiment studies. The compatibility with the 3GPP communication channel model and how to use this framework for sensing evaluation are also demonstrated.
\end{abstract}


\IEEEpeerreviewmaketitle

\section{Introduction}

\IEEEPARstart{I}{ntegrated} sensing and communication (ISAC) is considered to be a highly promising technology in the beyond 5G (B5G) and 6G wireless communication network \cite{survey}. 
For both communication and sensing functions, channel modeling is crucial to system design and evaluation. Thus, the standardization of a new technique is usually started with the channel modeling, so as the ISAC system.  
Therefore, channel modeling for ISAC has gained great attention in both academia and industry in recent years.

There exist many channel models for communication systems such as \cite{METIS, 36873,38901}. Particularly, the 3rd Generation Partnership Project (3GPP) TR 38.901 \cite{38901} provides a geometry-based stochastic channel model at the frequency from 0.5 GHz to 100 GHz for 5G systems. These models are primarily designed for communication systems. When wireless sensing is further considered, new channel models are required for ISAC channels. Currently, ISAC has already been considered in 3GPP standard with feasibility study \cite{22837} and service requirements \cite{22137}. More specifically, 3GPP TR 22.837 \cite{22837} provides 32 potential use cases of wireless sensing, which are classified into three categories in 3GPP TS 22.137 \cite{22137} based on the performance requirements of sensing, including object detection and tracking, environment monitoring, and motion monitoring. Furthermore, in many current ISAC studies, e.g., \cite{kumari2017ieee,zhao2023reference}, the communication channel and sensing channel are constructed separately, in which the correlation between two channels is not considered. However, to accurately evaluate the performance of an ISAC system, an ISAC channel should be utilized. Therefore, it is urgent to investigate the ISAC channel modeling.

\begin{figure}[t]
    \centering
    \subfigure[Monostatic sensing]{
	\includegraphics[width=0.45\linewidth]{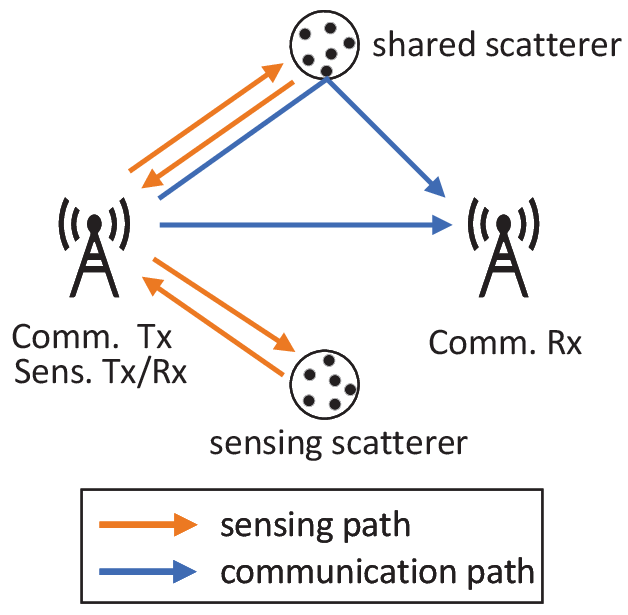}
	\label{fig:monostatic}
    }
    \subfigure[Bistatic sensing]{
        \includegraphics[width=0.45\linewidth]{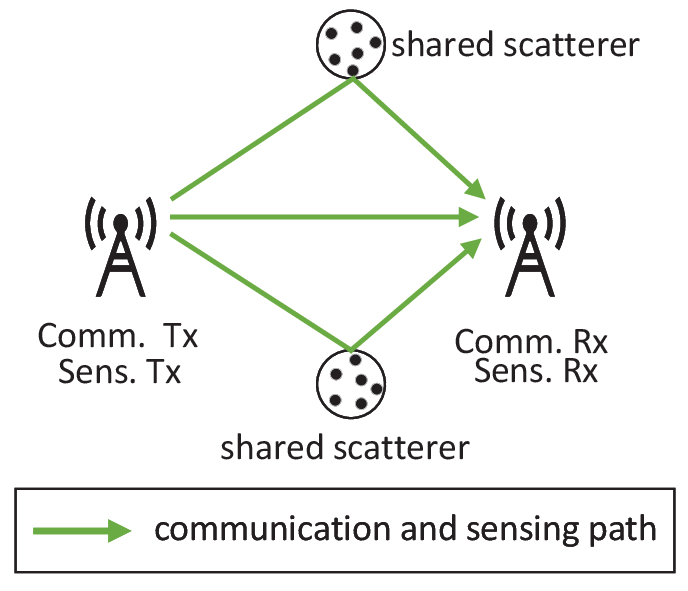}
        \label{fig:bistatic}
    }
    \caption{Diagram of the two sensing modes in the ISAC system.}
    \vspace{-0em}
\end{figure}

In order to develop the channel model for ISAC, a preliminary understanding of wireless sensing modes is essential. There are two basic wireless sensing modes, i.e., monostatic sensing and bistatic sensing. In monostatic sensing, the transmitter and the sensing receiver are the same transceiver, as is shown in Fig.\ref{fig:monostatic}. The sensing paths are from the target reflection back to the same transceiver, while the communication paths are the communication link from the transmitter to another communication receiver. In contrast, in bistatic sensing, the communication receiver and sensing receiver are the same node, which is separated from the transmitter, as is illustrated in Fig. \ref{fig:bistatic}. The communication and sensing paths all reach the same receiver. Due to the significant difference between the two sensing modes, the ISAC channel modeling for them should be separately investigated.

Currently, the channel modeling for ISAC system has been investigated by many studies, some of which are focused on monostatic channel modeling \cite{isacchannel3,yang2023novel,isacchannel1,zhang2024cluster,zhang2023shared,isacchannel2,liu2023shared,liu2024extend,xiong2023channel} and some of which are designed for both monostatic and bistatic channel modeling \cite{isacchannel4,yang2024integrated}. In \cite{isacchannel3}, the monostatic ISAC channel modeling is considered. More specifically, scatterers are divided into forward scatterers and backward scatterers, and a geometry-based stochastic channel model is then given in the sense that all paths in the monostatic sensing channel are from backward scattering, while the communication channel consists of backward scattering components from sensing. Moreover, the communication receiver can also be modeled as a sensing target. A similar approach is also applied in \cite{yang2023novel}, which uses monostatic backscattering sensing to assist the construction of communication channel. However, these channel modeling approaches do not follow the 3GPP framework \cite{38901}. In order to make the ISAC channel modeling compatible with the standard 3GPP framework, many studies explore the extended framework for ISAC channel modeling. In \cite{isacchannel1}, an extended framework is proposed, where communication and sensing share some clusters, and the birth and death process from communication clusters to sensing clusters is presented in the channel modeling. Followed by \cite{isacchannel1}, the authors further extend the framework to support the dedicated bistatic sensing \cite{zhang2024cluster}, where the sensing receiver is a dedicated sensing radar (not the communication receiver). The detail of the construction of shared clusters is further discussed in \cite{zhang2023shared}. The key feature of shared clusters in monostatic ISAC system is also investigated by \cite{isacchannel2,liu2023shared}, where experiment measurements are conducted to verify the cluster sharing between the communication channel and monostatic sensing channel. The concept of sharing degree is further introduced to characterize the sharing clusters. Furthermore, the extended stochastic channel modeling framework under 3GPP is further proposed considering the sharing feature \cite{liu2024extend}. In \cite{xiong2023channel}, the method of deterministic modeling of sensing targets and statistical modeling of communication clusters is adopted, considering shared clusters.
Besides the studies of channel modeling for monostatic ISAC, the extended framework under 3GPP standard that supports both monostatic and bistatic sensing has also gained great interest.
In \cite{isacchannel4}, a unified model is proposed for both monostatic and bistatic sensing modes, which extends the 3GPP framework to ISAC channel modeling. In this extended framework, the background clusters are all generated from the current 3GPP procedure, and some sensing targets are added as new clusters in the ISAC channel. A similar approach is also adopted in \cite{yang2024integrated}. Different from the statistical modeling of sensing targets in \cite{isacchannel4}, a deterministic approach is adopted for the sensing targets in \cite{yang2024integrated}. Thus, the framework in \cite{yang2024integrated} is a hybrid modeling approach for ISAC. More specifically, a deterministic multi-scattering-center (MSC) modeling of sensing targets is proposed for the sensing channel. The MSC extraction and modeling are further discussed in \cite{chen2024multi}. However, as for the bistatic case, the ISAC channel generated by the framework of \cite{isacchannel4} and \cite{yang2024integrated} is not fully compatible with the pure communication channel, since more sensing clusters are directly added to the communication channel. Moreover, such an approach cannot well reflect the weak environment clutters for sensing, since the communication channel modeling in the 3GPP standard only retains relatively strong clusters.
Thus, how to model the bistatic ISAC channel remains an open research problem. A summary of the main ISAC channel modeling approaches is presented in Table \ref{tab:summary}.

\begin{table*}[t]
    \centering
    \caption{A summary of the main ISAC channel modeling approaches}
    \label{tab:summary}
    \begin{tabular}{ccccc}
    \toprule
    Reference  & ISAC scenario & \shortstack{extension of \\ 3GPP framework} & \shortstack{modeling\\ method} & correlation between communication and sensing\\
    \midrule
    \cite{yang2023novel}  & monostatic & no & statistical & \shortstack{communication channel consists of backward \\scattering components from sensing} \\
    \hline
    \cite{isacchannel1} &  monostatic & yes & statistical & \shortstack{communication and sensing share some clusters; \\ birth and death process is proposed for communication clusters to sensing clusters}  \\
    \hline
    \cite{liu2024extend} & monostatic & yes & statistical & \shortstack{communication and sensing share some clusters; \\ shared degree is defined and introduced}\\
    \hline
    \cite{isacchannel4} & \shortstack{monostatic and \\ bistatic} & yes & statistical & \shortstack{communication channel is utilized as target unrelated environment channel; \\targeted related sensing channel is  separately generated and added}\\
    \hline
    \cite{yang2024integrated} & \shortstack{monostatic and \\ bistatic} & yes& hybrid &  \shortstack{communication channel is utilized as target unrelated environment channel; \\ deterministic MSC model is used to add sensing targets} \\
    \hline
    this paper & bistatic & yes & \shortstack{statistical \\or hybrid} & \shortstack{generate more communication clusters following 3GPP framework; \\convert some of them to sensing clusters according to sensing scenario}\\

    \bottomrule
    \end{tabular}
    \vspace{-0em}
\end{table*}

In this paper, the channel modeling framework for bistatic ISAC channel is developed under the current standard framework of 3GPP TR 38.901 \cite{38901}. Unlike adding independent sensing targets into the communication channel in \cite{isacchannel4,yang2024integrated}, the sensing targets in our proposed modeling approach are first generated under the 3GPP framework and then some of them are converted to sensing clusters according to sensing scenarios. More specifically, to support sensing function, we generate more clusters following the 3GPP framework, where more weak power clusters are retained to reflect the fact that radar sensing can detect much weaker targets. These retained weak clusters can also reflect sensing clutters. After that, some clusters are selected as sensing clusters according to sensing scenarios. For sensing clusters, either statistical or deterministic modeling method can be applied to get the sensing rays, depending on the sensing scenarios. Moreover, regardless of which method is adopted, the time and spatial coherence are ensured for sensing clusters. A unified channel is finally generated for both communication and sensing. Therefore, the generated ISAC channel can be used to evaluate both communication function and sensing function.

The contributions of this paper are summarized as follows.
\begin{enumerate}
    \item A bistatic ISAC channel modeling framework is proposed in this paper, which extends the current 3GPP channel modeling framework \cite{38901} for both communications and bistatic sensing. In the extended modeling framework, a few communication clusters are converted to sensing clusters, with either statistical or deterministic modeling method. The proposed ISAC channel is of high compatibility with the current 3GPP communication channel model.
    \item Several key features for bistatic ISAC channel modeling are revealed, including: 1) the communication scenarios and sensing scenarios should be jointly considered; 2) more weak clusters should be included in the bistatic ISAC channel than that in communication channel; 3) both spatial and time coherence should be ensured for sensing clusters, either in statistical or deterministic sensing target modeling.
    \item Extensive ray tracing simulations and experiment measurements are carried out to validate the effectiveness of our proposed framework. The key features for bistatic sensing channel requirements are verified. Furthermore, the compatibility with the 3GPP communication channel model and how to use this framework for sensing evaluation are also demonstrated.
\end{enumerate}

The rest of the  paper is outlined as follows. In Section \uppercase\expandafter{\romannumeral2}, the framework and detailed modeling steps of the proposed bistatic ISAC channel model are elaborated. Simulation and experiment validation of our proposed channel model are carried out in \uppercase\expandafter{\romannumeral3}. This paper is concluded in \uppercase\expandafter{\romannumeral4}.

\section{Bistatic ISAC Channel Modeling}

In this section, the bistatic ISAC channel modeling framework is proposed. The design insight is first presented, and then the detailed steps under the framework are illustrated.

\subsection{Design Insight}

In the bistatic ISAC system, since the physical propagation channel is identical for the communication function and the sensing function, the communication channel and the sensing channel should share the same channel parameters. However, the current channel modeling scheme in the 3GPP standard \cite{38901} cannot be directly applied to the sensing function, due to the following two-fold reasons. First, paths with very low energy in \cite{38901} are omitted due to their little contributions to communication performance. However, these low energy paths may be crucial for target sensing. In radar sensing processing, a large radar coherent processing gain can be achieved to combat noise. Thus, wireless sensing can be achieved under a very low signal-to-noise (SNR) ratio such as $-20$ dB \cite{berger2010signal}. Therefore, these paths should be retained in the ISAC channel. Second, the time-and-spatial coherence in \cite{38901} is not well addressed, since its impact on communication performance is not significant. However, the time-and-spatial coherence is critical for sensing applications. Therefore, more detailed information for target-related paths is required, and the modeling method should be sensing application-oriented. For example, behavior recognition requires fine-grained parameters of the propagation rays, which can be characterized by deterministic modeling; while for applications like target localization and tracking, statistical modeling with time-and-spatial coherence is enough to contain the required sensing information.


Based on the above insights, more clusters with weak power are generated and retained in the proposed bistatic ISAC channel modeling scheme. Furthermore, the environment-related paths and target-related paths are separately modeled, where the former can follow the standard channel modeling for communications, and the latter needs to carry more fine-grained information with well time-and-spatial coherence. Moreover, the modeling method for target-related paths depends on the sensing applications.



\begin{figure*}[t]
    \centering
    \includegraphics[width=0.99\linewidth]{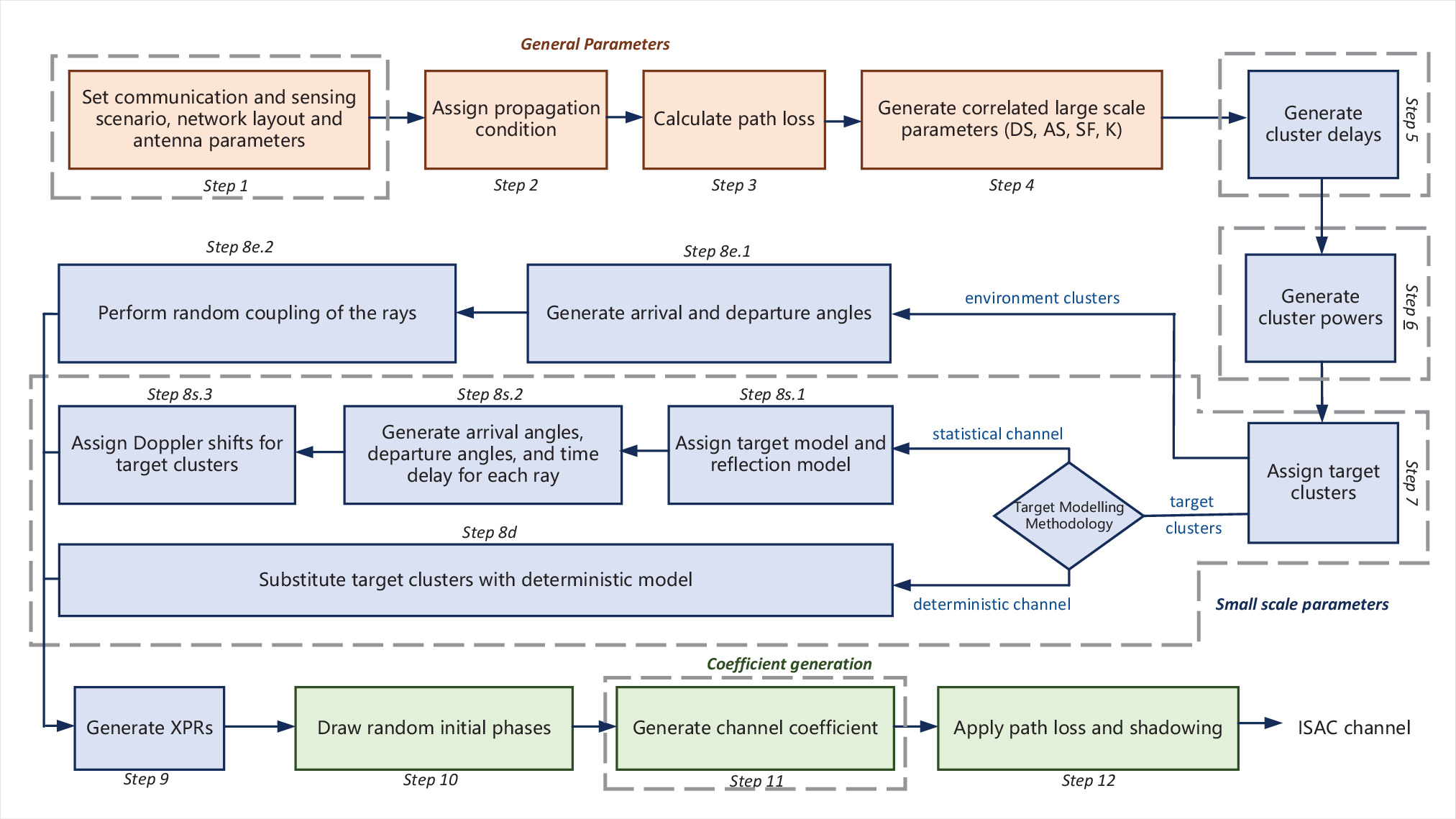}
    \caption{Framework of bistatic ISAC channel modeling.}
    \label{fig:frame}
    \vspace{-0em}
\end{figure*}

To support wireless sensing, more features should be contained in the bistatic ISAC channel model. However, such an ISAC model should have high compatibility with the current communication models. Therefore, to ensure compatibility with existing communication models, the proposed framework is extended from 3GPP TR 38.901\cite{38901}. The diagram of the framework is shown in Fig.~\ref{fig:frame}. The procedure of channel modeling is categorized into three parts (general parameter generation, small scale parameter generation, and coefficient generation), where the modification of the 3GPP channel model is highlighted by dash boxes. The output channel coefficients of the ISAC channel are suitable for both communication and sensing purposes due to the identical physical environment for both communications and sensing.

\begin{table}[t]
    \centering
    \renewcommand{\arraystretch}{1.2}
    \renewcommand{\tabcolsep}{1.2mm}
    \caption{Examples of ISAC scenarios}
    \label{tab:scenario}
    \begin{tabular}{ll}
    Communication scenario  & Sensing scenario\\
    \hline
    urban micro (UMi)  & target localization \\
    urban macro (UMa) &  behavior recognition \\
    indoor office & breath detection \\
    rural macro (RMa) & invasion detection  \\
    indoor factory (InF) & environment imaging\\
    \hline
    \end{tabular}
    \vspace{-0em}
\end{table}

\subsection{ISAC Channel Modeling Framework}
\subsubsection{General Parameters Generation}
As is shown in Fig. \ref{fig:frame}, the generation of general parameters is from step 1 to step 4. In the 3GPP model, a communication scenario such as urban micro (UMi) and urban macro (UMa) is assigned at the first step. For the ISAC channel, in addition to the communication scenario, the sensing scenario is further assigned. A communication scenario and a sensing scenario together form an ISAC scenario.
Some typical examples of sensing scenarios are shown in Table \ref{tab:scenario}, such as target localization and behavior recognition. The determination of sensing scenario affects the following target model and related sensing cluster parameter generation. In short, ISAC channel modeling requires the input of both a communication scenario and a sensing scenario.
Other general parameters in step 1 are generated following the 3GPP procedure \cite{38901}. 
The following step 2 to step 4 can also directly follow the standard procedure in \cite{38901} to generate necessary general parameters.


\subsubsection{Small Scale Parameters Generation}
As is shown in Fig. \ref{fig:frame}, the generation of small scale parameters is from step 5 to step 9.
In Step 5, clusters are generated with different delays.
To support wireless sensing for weak targets, more clusters are required for an ISAC channel, compared to that for a communication channel. Thus, the number of clusters $N$ in the standard procedure \cite{38901} is modified to be a larger value $N_{\text{ISAC}}$, so that more weak clusters can be generated.
For each cluster $n$, the generation of cluster delay $\tau_n$ still follows the standard procedure \cite{38901} as
\begin{equation}
    \tau_n^{\prime}=-r_\tau~\mathrm{DS}~\mathrm{ln}(X_n),
\end{equation}
and is normalized and sorted  to ascending order by
\begin{equation}
    \tau_n=\operatorname{sort}(\tau_n^{\prime}-\min(\tau_n^{\prime})),
\end{equation}
where $r_\tau$ and $\text{DS}$ denote delay distribution proportionality factor and delay spread respectively, which are adopted from the large scale parameters, and $X_n \sim \text{uniform}(0,1)$ is the random variable.

The generation of cluster power in step 6 also follows the standard procedure \cite{38901} by
\begin{equation}
    P_n'=\exp\Bigg(-\tau_n\frac{r_\tau-1}{r_\tau\text{DS}}\Bigg ) \cdotp 1 0 ^ { \frac { - Z_n}{10}},
\end{equation}
which is further normalized by
\begin{equation}\label{eq:power}
    P_n=\frac{P_n^{\prime}}{\sum_{n=1}^{N_{\text{ISAC}}}P_n^{\prime}}
\end{equation}
where $Z_{n}\sim N{\left(0,\zeta^{2}\right)}$ is the random variable indicating shadowing term per cluster.

With the generated power in Eq. (\ref{eq:power}), some weak clusters are removed from the channel model based on a certain threshold. In 3GPP standard\cite{38901}, the threshold is set to be $-25$ dB, which is defined as the cluster with power more than 25 dB lower than the cluster with maximum power.
For the communication function, the contribution to channel gain of these weak clusters can be ignored. However, for sensing function, some of these removed weak clusters can still be sensed. Therefore, a \emph{much lower threshold} for removing clusters is required in ISAC channel modeling. The threshold can be sensing scenario-dependent. For scenarios focused on sensing targets with relatively strong signal strength, such as respiration detection or behavior recognition, a higher threshold can be used. Conversely, for scenarios involving sensing targets with lower signal strength and significant environmental clutter impacts, such as outdoor localization, a lower threshold should be adopted.
The exact threshold for each scenario needs to be explored in the further study under this framework.  

In step 7, the retained clusters are classified into target clusters and environment clusters. In other words, some clusters are selected and converted to sensing clusters. Since the power and delay for each cluster have been assigned in previous steps, the clusters for sensing should be properly selected. Some guidelines for the selection are suggested. First, the selection of sensing clusters should depend on the application scenarios defined in the first step. For example, the number of sensing targets may already be specified, such as single-target or multi-target sensing. Second, given a sensing scenario, the selection of sensing clusters is further determined to satisfy the sensing requirement. Some examples are as follows. For outdoor radar sensing, the received power of a target is related to both the range and radar cross section (RCS). Different targets usually have different RCSs. For example, the typical value of RCS for a man and automobile is 0 dBsm and 20 dBsm, respectively. Therefore, the selection of the sensing cluster needs to jointly consider the effect of target type in the sensing scenario, delay, and power. In scenarios such as target location in an indoor office or invasion detection, the detection region is a key parameter. The selection of the target cluster can be based on the defined detection region and clusters within a certain range of delay are selected. For other applications like respiration detection or behavior recognition, the reflection of target is usually considered as a strong NLoS path in order to capture its Doppler pattern of motion. In these scenarios, the cluster with the second strongest cluster power may be selected to meet the requirements of sensing applications. In short, different sensing applications have varying sensing parameters, and the selection of target clusters should be guided by the specific requirements of these parameters.
The non-selected clusters are environment clusters. Since more weak clusters are retained, these non-selected weak clusters can also be treated as the weak environment clutters for sensing.




For the environment clusters, the 3GPP standard procedure \cite{38901} is used to generate the arrival and departure angles for each cluster and perform random coupling of rays in step 8e.1 and 8e.2, respectively. 
For the target clusters, either statistical modeling or deterministic modeling can be applied to generate rays for each sensing cluster, based on the ISAC scenario. Statistical modeling is suitable for sensing applications where the target exhibits a regular pattern of motion, e.g. breath detection and target localization, or there is no high requirement of the sensing result, e.g. invasion detection; while deterministic modeling is suitable for the application where fine-grained information of the rays is required, e.g. behavior recognition and environment imaging.

\begin{figure}[t]
    \centering
    \includegraphics[width=0.8\linewidth]{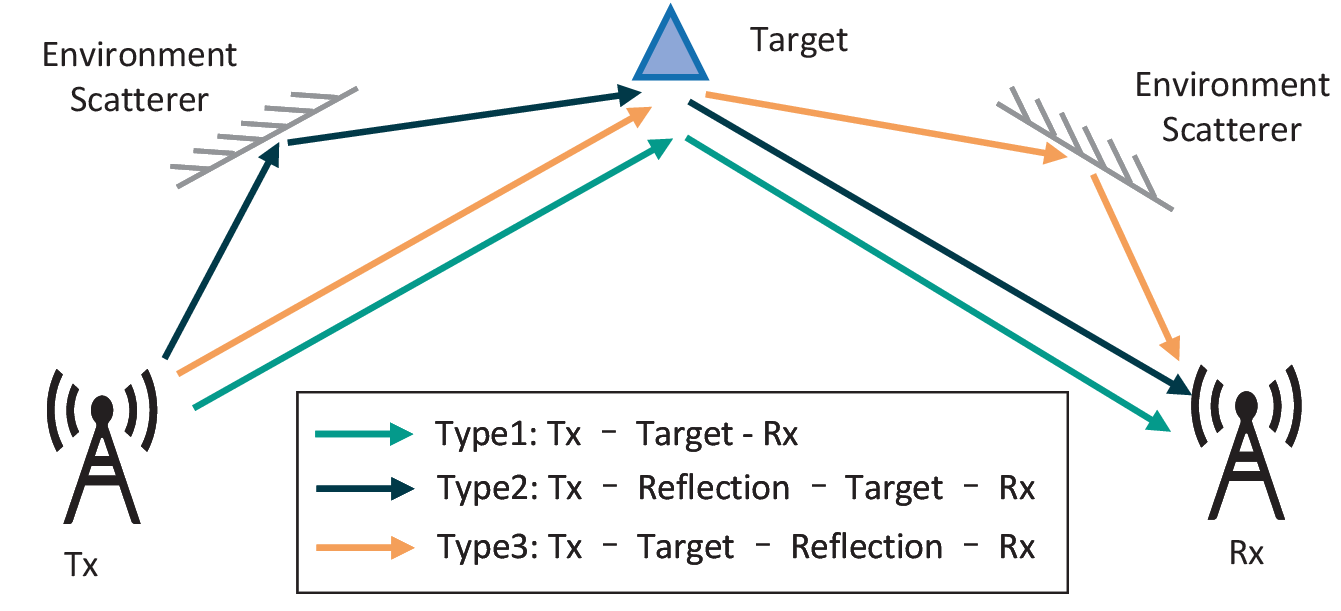}
    \caption{Three types of reflection model for sensing rays in a target cluster.}
    \label{fig:reflection}
    \vspace{-0em}
\end{figure}

For statistical modeling, the generation of target cluster first assigns the target model and reflection model in Step 8s.1. Target model can be either point target model or extended target models, which is determined by the ISAC scenario. For example, for the scenario of urban macro (UMa) for communications and target localization for sensing, the point target model is enough for the evaluation of sensing performance; while for the scenario of indoor office for communications and target localization, extended target model may be required.
The reflection model determines the bounce of sensing rays. To ensure the sensing capability in the ISAC channel model, the sensing target should be the first scatterer or the last scatterer. To get the position of the sensing target, it should be both first and last scatter. If the sensing target is the first scatter, we can extract the azimuth angle of departure (AOD) and zenith angle of departure (ZOD) information (or possible relative departure velocity) from the sensing channel. Similarly, if the sensing target is the last scatter, we can extract the azimuth angle of arrival (AOA) and zenith angle of arrival (ZOA) information (or possible relative arrival velocity) from the sensing channel. However, if the sensing target is neither the first scatter nor the last scatter, we can hardly extract sensing information from the sensing channel. Moreover, since the signal strength is dramatically reduced for each bounce, the total number of bounces for sensing rays should be limited. For example, each ray can be considered to have at most 2 bounces, including the environment reflection and the target reflection. As illustrated in Fig. \ref{fig:reflection}, three types of reflection models for target cluster are considered: 1) Tx - Target - Rx; 2) Tx - Reflection - Target - Rx; 3) Tx - Target - Reflection - Rx. The sensing cluster can include multiple reflection types at the same time.
The determination of the reflection model for each ISAC scenario needs to be explored in the further work under this framework.

When the target model and reflection model are determined, the target cluster can be further modeled as one cluster or a few sub-clusters. For example, if point target model is applied, single ray is applied for the target. However, if one type 1 reflection and one type 2 reflection are further applied, the target cluster can include two sub-clusters, each with one ray inside. If extended target model is applied, multiple rays are required for each reflection. If multiple bounces are further considered, the target cluster is modeled by a few sub-clusters, each with multiple rays inside. The generated cluster power can be divided into all the sub-clusters with a power distribution related to the ISAC scenario, which needs to be modeled in further work. 

Next, to guarantee the sensing capability of the ISAC channel model, the spatial coherence for the angle and delay is ensured in step 8s.2 as follows. Based on the target model and reflection model, one sensing cluster may have a few sub-clusters.
The parameters of time delay ($\tau_{n,m}$), arrival angles ($\theta_{n,m,\text{ZOA}}$, $\phi_{n,m,\text{AOA}}$), and departure angles ($\theta_{n,m,\text{ZOD}}$, $\phi_{n,m,\text{AOD}}$) for the m-th ray of the n-th target cluster are generated following a geometric-based manner with spatial coherence. Two options (angle-priority and position-priority) are provided here for generating these parameters.

In the angle-priority method, the angle information is used for the initial target position generation. Use the departure angle as an example, the procedures are illustrated as follows:

\begin{enumerate}
    \item Generate departure angles $\theta_{n,\text{ZOD}}$, $\phi_{n,\text{AOD}}$ for each target cluster $n$ following the 3GPP procedure.
    \item Calculate the 3D location $\text{\textbf{R}}_{n}$ of the equivalent reflection point of the cluster $n$. $\text{\textbf{R}}_{n}$ is the intersection between the ray emitted from the transmitter of the direction of $(\theta_{n,\text{ZOD}}, \phi_{n,\text{AOD}})$ and the ellipse with Tx and Rx as the focus and $\tau_n\cdot c$ as the major axis ($c$ is the speed of light).
    \item Parameters for type 1 rays are first generated. For point target model, $\text{\textbf{R}}_{n}$ is treated as the only reflection point for the target; for extended target model, multiple reflection points $\text{\textbf{R}}_{n,m}$ are randomly generated around the cluster central reflection point $\text{\textbf{R}}_{n}$ with an angle spread factor. Each reflection point $\text{\textbf{R}}_{n,m}$ leads to one ray. For each ray, calculate its time delay $\tau_{n,m}$, departure angles ($\theta_{n,m, \text{ZOD}}$, $\phi_{n,m,\text{AOD}}$), and arrival angles ($\theta_{n, m, \text{ZOA}}$, $\phi_{n, m, \text{AOA}}$) based on the 3D location of $\text{\textbf{R}}_{n,m}$. The total number of rays and angle spread factor depend on a specific extended target model, which can be further defined under this framework.
    \item Type 2 rays are extended from type 1 rays. In type 2 reflection model, the target is the second scatterer, and its reflection points  $\text{\textbf{R}}_{n,m}$ for the second bounce are inherited from the type 1 ray. The first reflection point $\text{\textbf{R}}_{n,m,\text{env}}$ of the environment scatterer is randomly generated for ray $(n,m)$. The time delay $\tau_{n,m}$ is determined by the path of Tx - $\text{\textbf{R}}_{n,m,\text{env}}$ - $\text{\textbf{R}}_{n,m}$ - Rx, and its departure angle is regenerated as the angle from transmitter to  $\text{\textbf{R}}_{n,m,\text{env}}$.
    \item The procedure for the generation of type 3 rays is similar to that for type 2 rays, except that the target is the first scatterer and the arrival angle for each ray is regenerated. 
\end{enumerate}

While in the position-priority method, the position of the reflection point is generated first. The procedures are illustrated as follows:

\begin{enumerate}
    \item For each cluster, randomly choose a point $\text{\textbf{R}}_{n}$ to be the equivalent reflection point on the ellipse with Tx and Rx as the focus and $\tau_n\cdot c$ as the major axis ($c$ is the speed of light).
    \item The remaining steps are the same as described in step 3) - step 5) in the angle-priority methods
\end{enumerate}

As we can see, the only difference between these two options is the generation of the initial 3D target location. The initial position is more random in option 2, while it is constrained by the angle information in option 1. The selection of the option depends on the sensing scenario.

In step 8.s3, Doppler shifts are assigned, which is based on the sensing scenario and time coherence.

If the target is modeled in a deterministic manner, as in step 8d, all the channel coefficients for the target clusters are replaced by the results of ray tracing simulation or real measurements, including power $P_{n,m}$, time delay $\tau_{n,m}$, arrival angles ($\theta_{n,m,\text{ZOA}}$, $\phi_{n,m,\text{AOA}}$), and departure angles ($\theta_{n,m,\text{ZOD}}$, $\phi_{n,m,\text{AOD}}$). 

The cross polarization power ratio (XPR) $\kappa$ for each ray is generated in step 9, which can follow the 3GPP standard procedure \cite{38901}.

\subsubsection{Coefficient Generation}
As is shown in Fig. \ref{fig:frame}, the coefficient generation includes the final three steps. Random initial phases $\Phi_{n, m}^{\theta \theta}$,  $\Phi_{n, m}^{\theta \phi}$, $\Phi_{n, m}^{\phi \theta}$, and $\Phi_{n, m}^{\phi \phi}$ are generated in step 10 for each ray of all clusters.

The channel coefficients are generated in step 11 to get the small-scale channel impulse response (CIR), where the sensing cluster requires additional design for time coherence. Basically, there are two cases: LoS case or NLoS case.
The ISAC NLoS channel impulse response between the $u$-th UT and $s$-th BS is given by
\begin{equation}
\begin{aligned}\label{eq:nlos}
    H_{u,s}^{\mathrm{NLOS}}(\tau, t)=& \sum_{n\in \mathcal{T}} \sum_{m=1}^{M_{n}} H_{u, s, n,m}^{\mathrm{NLOS, tar}}(t) \delta\left(\tau-\tau_{n,m}\right) \\
    & +\sum_{n\in \mathcal{E}}\sum_{m=1}^{M} H_{u, s, n, m}^{\mathrm{NLOS, env}}(t) \delta\left(\tau-\tau_{n,m}\right),
\end{aligned}
\end{equation}
where $\mathcal{T}$ and $\mathcal{E}$ denote the cluster set for sensing targets and environment clusters, respectively; $M_n$ represents the number of rays in the $n$-th target cluster; $M$ denotes the number of rays in the environment cluster, whose value is specified in the 3GPP standard.
The generation of $H_{u, s, n, m}^{\mathrm{NLOS, env}}(t)$ can directly follows the formula in 3GPP standard \cite{38901} as follows.
\begin{equation}
\begin{array}{l}H_{u, s, n, m}^{\mathrm{NLOS,env}}(t)=\sqrt{\frac{P_{n}}{M}}\left[\begin{array}{l}F_{\text{rx}, u, \theta}\left(\theta_{n, m, \text{ZOA}}, \phi_{n, m, \text{AOA}}\right) \\
F_{\text{rx}, u, \phi}\left(\theta_{n, m, \text{ZOA}}, \phi_{n, m, \text{AOA}}\right)\end{array}\right]^{T}\\\left[\begin{array}{cc}\exp \left(j \Phi_{n, m}^{\theta \theta}\right) & \sqrt{\kappa_{n, m}{ }^{-1}} \exp \left(j \Phi_{n, m}^{\theta \phi}\right) \\
\sqrt{\kappa_{n, m}{ }^{-1}} \exp \left(j \Phi_{n, m}^{\phi \theta}\right) & \exp \left(j \Phi_{n, m}^{\phi \phi}\right)\end{array}\right] \\ \left[\begin{array}{l}F_{\text{tx}, s, \theta}\left(\theta_{n, m, \text{ZOD}}, \phi_{n, m, \text{AOD}}\right) \\
F_{\text{tx}, s, \phi}\left(\theta_{n, m, \text{ZOD}}, \phi_{n, m, \text{AOD}}\right)\end{array}\right] \exp \left(j 2 \pi \frac{\hat{r}_{\text{rx}, n, m}^{T} \cdot \bar{d}_{\text{rx}, u}}{\lambda_{0}}\right)\\
\exp \left(j 2 \pi \frac{\hat{r}_{\text{tx}, n, m}^{T} \cdot \bar{d}_{\text{tx}, s}}{\lambda_{0}}\right) \exp \left(j 2 \pi \frac{\hat{r}_{\text{rx}, n, m}^{T} \cdot \bar{v}_{\text{UT}}}{\lambda_{0}} t\right),
\end{array}
\label{eq:NLOS}
\end{equation}
where $F_{\text{rx}, u, \theta}$, $F_{\text{rx}, u, \phi}$, $F_{\text{tx}, s, \theta}$, and $F_{\text{tx}, s, \phi}$ are the field patterns of the $u$-th UT and $s$-th BS; $\hat{r}_{\text{rx}, n,m}$ and $\hat{r}_{\text{tx}, n,m}$ denote the spherical unit vector with its corresponding azimuth arrival angle and elevation arrival angle, while $\bar{d}_{\text{rx}, u}$ and $\bar{d}_{\text{tx}, s}$ denote the location vector at the antennas of UT and BS respectively; $\lambda_0 $ is the wavelength of carrier frequency; $\bar{v}_{\text{UT}}$ is the time-invariant velocity of UT, as is defined in the 3GPP standard. The total power is $P_n$.

The generation of $H_{u, s, n,m}^{\mathrm{NLOS, tar}}(t)$ depends on the used modeling method for sensing cluster.
For statistically modeled sensing cluster, the channel coefficient $H_{u, s, n,m}^{\mathrm{NLOS, tar}}(t)$ is given by:
\begin{equation}
\begin{array}{l}H_{u, s, n, m }^{\mathrm{NLOS, tar}}(t)=\sqrt{P_{n,m}}\left[\begin{array}{l}F_{\text{rx}, u, \theta}\left(\theta_{n,m, \text{ZOA}}, \phi_{n, m, \text{AOA}}\right) \\
F_{\text{rx}, u, \phi}\left(\theta_{n, m, \text{ZOA}}, \phi_{n,m, \text{AOA}}\right)\end{array}\right]^{T}\\\left[\begin{array}{cc} \exp \left(j \Phi_{n,m}^{\theta \theta}\right) & \sqrt{\kappa_{n,m}{ }^{-1}} \exp \left(j \Phi_{n, m}^{\theta \phi}\right) \\
\sqrt{\kappa_{n,m}{ }^{-1}} \exp \left(j \Phi_{n,m}^{\phi \theta}\right) & \exp \left(j \Phi_{n,m}^{\phi \phi}\right)\end{array}\right] \\ \left[\begin{array}{l}F_{\text{tx}, s, \theta}\left(\theta_{n,m, \text{ZOD}}, \phi_{n,m, \text{AOD}}\right) \\
F_{\text{tx}, s, \phi}\left(\theta_{n, m, \text{ZOD}}, \phi_{n,m, \text{AOD}}\right)\end{array}\right] \exp \left(j 2 \pi \frac{\hat{r}_{\text{rx}, n,m}^{T} \cdot \bar{d}_{\text{rx}, u}}{\lambda_{0}}\right)\\
\exp \left(j 2 \pi \frac{\hat{r}_{\text{tx}, n, m}^{T} \cdot \bar{d}_{\text{tx}, s}}{\lambda_{0}}\right) \exp \left(j 2 \pi \frac{\hat{r}_{\text{rx}, n,m}^{T} \cdot \bar{v}_{\text{UT}}}{\lambda_{0}} t\right)\\
\exp\left(j2\pi\int_{0}^t\frac{v_{n, m}\left(\widetilde{t}\right)}{\lambda_0} d\tilde{t}\right),
\end{array}
\end{equation}
where $P_{n,m}$ is the power of the $m$-th ray of the $n$-th target cluster, depending on the cluster model. 
Compared to the environment cluster, just one more term $\exp\left(j2\pi\int_{0}^t\frac{v_{n, m}(\widetilde{t})}{\lambda_0} d\tilde{t}\right)$ related to sensing target is added. In order to keep the time coherence for sensing target, a new factor $v_{n,m}(t)$ is introduced as the effective velocity of the target at time $t$, which will result in an additional phase shift due to Doppler Shift. The integral of $\exp\left(j2\pi\int_{0}^t\frac{v_{n, m}(\widetilde{t})}{\lambda_0} d\tilde{t}\right)$ represents the cumulative phase shift by Doppler shift from the beginning of the simulation to the certain timestamp $t$. With the setup of velocity for each sensing target, the integral term ensures the time coherence for sensing. 
To maintain time coherence for sensing, the entire process of generating clusters is conducted only once. During one radar coherent processing interval (CPI), the delay and effective velocity for each target keep constant. However, as time progresses, the position of each target changes and the velocity of each target can also vary. Thus, the corresponding delay and angle should be adjusted according to the updated position of each sensing target.

For deterministically modeled target cluster, phase shift and channel gain from time delay, beam pattern, polarization, antenna arrangement, initial phase, and Doppler shift have already been jointly contained in the ray tracing simulation results or measurement results as the phase of the path $\varphi_{n,m}$. The time coherence of the sensing cluster is guaranteed by the ray tracing process or the real-world experiment measurement.
Thus, the channel coefficient for target cluster is given by
\begin{equation}
    H_{u, s, n, m }^{\mathrm{NLOS, tar}}(t)=\sqrt{P_{n,m}^\prime}\exp \left( j  \varphi_{n,m} \right),
\end{equation}
where $P^\prime_{n,m}$ is the power gain value from the ray tracing simulation divided by the calculated path loss in step 1, i.e.,
\begin{equation}
    P_{n,m}^\prime = P_{n,m}/PL.
\end{equation}

In the LoS condition, the LoS path channel coefficient is given by the 3GPP standard \cite{38901}

\begin{equation}
\begin{array}{l}H_{u, s,1}^{\mathrm{LOS}}(t)=\left[\begin{array}{l}F_{\text{rx}, u, \theta}\left(\theta_{\text{LoS}, \text{ZOA}}, \phi_{\text{LoS}, \text{AOA}}\right) \\
F_{\text{rx}, u, \phi}\left(\theta_{\text{LoS}, \text{ZOA}}, \phi_{\text{LoS}, \text{AOA}}\right)\end{array}\right]^{T}
\left[\begin{array}{cc} 1 & 0 \\ 0 & -1\end{array}\right] \\ \left[\begin{array}{l}F_{\text{tx}, s, \theta}\left(\theta_{\text{LoS}, \text{ZOD}}, \phi_{\text{LoS}, \text{AOD}}\right) \\
F_{\text{tx}, s, \phi}\left(\theta_{\text{LoS}, \text{ZOD}}, \phi_{\text{LoS}, \text{AOD}}\right)\end{array}\right] \exp \left(j 2 \pi \frac{\hat{r}_{\text{rx}, \text{LoS}}^{T} \cdot \bar{d}_{\text{rx}, u}}{\lambda_{0}}\right)\\
\exp \left(j 2 \pi \frac{\hat{r}_{\text{tx}, \text{LoS}}^{T} \cdot \bar{d}_{\text{tx}, s}}{\lambda_{0}}\right) \exp \left(j 2 \pi \frac{\hat{r}_{\text{rx}, \text{LoS}}^{T} \cdot \bar{v}_{\text{UT}}}{\lambda_{0}} t\right) \exp \left(-j 2 \pi \frac{d_{\text{3D}}}{\lambda_{0}}\right),
\end{array}
\end{equation}
where $d_{\text{3D}}$ is the 3D distance between the BS and UT. The LoS channel impulse response is the sum of the LOS and NLoS channel coefficients with the scaling based on K-factor $K_R$, given by:
\begin{equation}
\begin{aligned}
H_{u,s}^{\mathrm{LOS}}(\tau,t)=&\sqrt{\frac1{K_R+1}}H_{u,s}^{\mathrm{NLOS}}(\tau,t)\\
    & +\sqrt{\frac{K_R}{K_R+1}}H_{u,s,1}^{\mathrm{LOS}}(t)\delta{\left(\tau-\tau_1\right)}.
\end{aligned}
\end{equation}

The path loss and shadow fading are applied to the above channel coefficients in step 12 to get the final bistatic ISAC channel. The above processing for generating channel coefficients can be applied to both cases of statistical modeling and deterministic modeling in step 8 for target clusters. However, there is also another choice for deterministic modeling case as follows. The Eq. (\ref{eq:nlos}) only calculates the effect of environment NLoS clusters, and the results of ray tracing or experiment results for sensing targets are directly summed into the final channel coefficients.

The generated bistatic ISAC channel can be used for both communications and sensing.

\section{Verification of the Bistatic ISAC Channel Model}

\subsection{Validation by Ray Tracing Simulations}
\subsubsection{Simulation setup} 
\begin{figure}[t]
    \centering
    \subfigure[The indoor office scenario]{
	\includegraphics[width=0.7\linewidth]{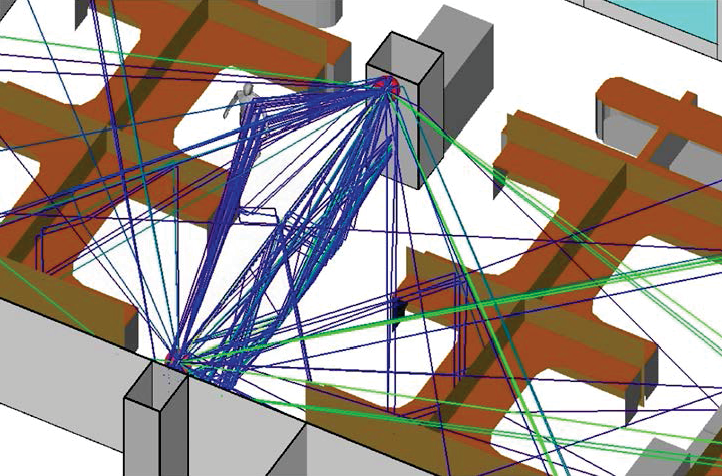}
    	\label{fig:office}
    }
    \subfigure[Channel impulse response]{
        \includegraphics[width=0.99\linewidth]{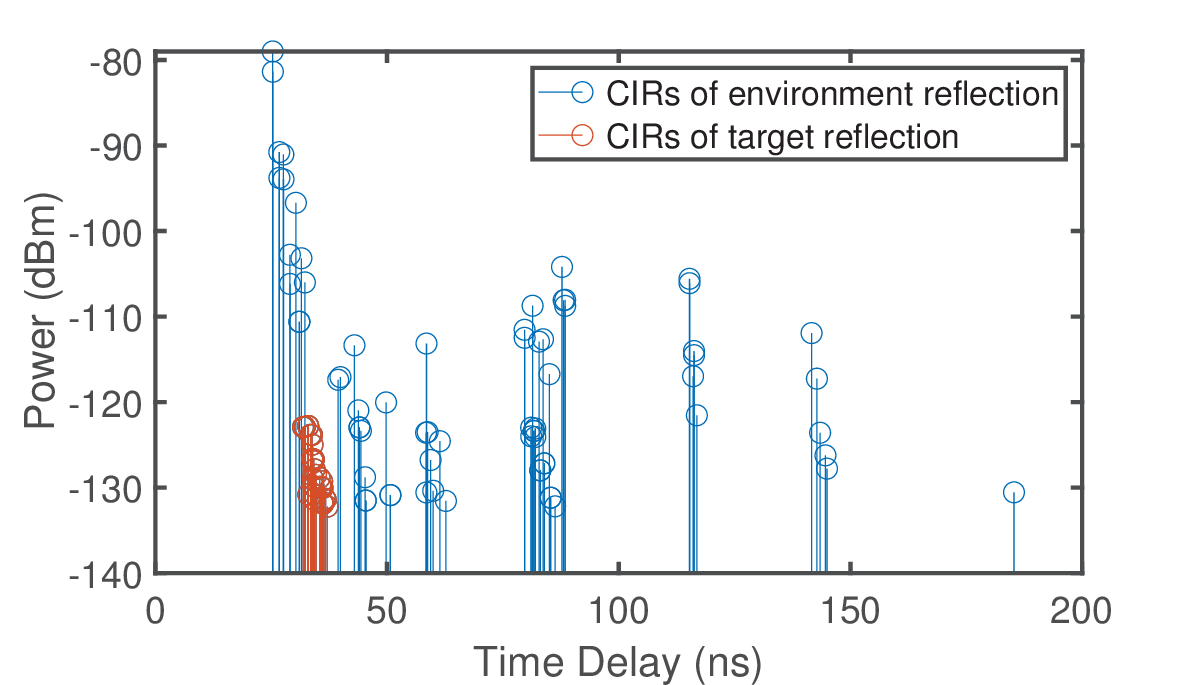}
        \label{fig:rayCIR}
    }
    \caption{Ray tracing simulation in an indoor office.}
    \vspace{-0em}
\end{figure}

Ray tracing is a method of simulating the behavior of light or electromagnetic waves as they propagate through a given environment by computing the Maxwell equations. Ray tracing can provide information regarding the propagation paths, attenuation, and angles of electromagnetic signals in the spatial domain. The ray tracing simulation is performed on the platform of Remcom Wireless Insite \cite{wirelessinsite}. The simulation scenario is set in an indoor office, with both the transmitter and the receiver placed on the wall and a human model as the target to be located by the desk, as is shown in Fig. \ref{fig:office}. The transmit antenna and the receive antenna are both isotropic antennas. The transmitter sends a single carrier waveform at 28 GHz with 500 MHz bandwidth and a transmission power of 0 dBm. The strongest 100 paths are collected by the ray tracing software.

\subsubsection{Power of target cluster}
The CIR for the environment set in Fig. \ref{fig:office} is illustrated in Fig. \ref{fig:rayCIR}. The power of LoS path is $-78.9$ dBm, while the rays related to the target cluster are within the power range from $-122$ dBm to $-132$ dBm. Other environment echos are within the power range from $-90$ dBm to $-134$ dBm. The result shows that the target clusters can have much less power than the power threshold ($-25$ dB) in the 3GPP model, which reveals the necessity of the generation of more clusters in step 5 and also a much lower power threshold to keep the weak sensing clusters in step 6. This result also shows the necessity of the extended target model for sensing in step 8s.

\begin{figure}[t]
    \centering
	\includegraphics[width=0.99\linewidth]{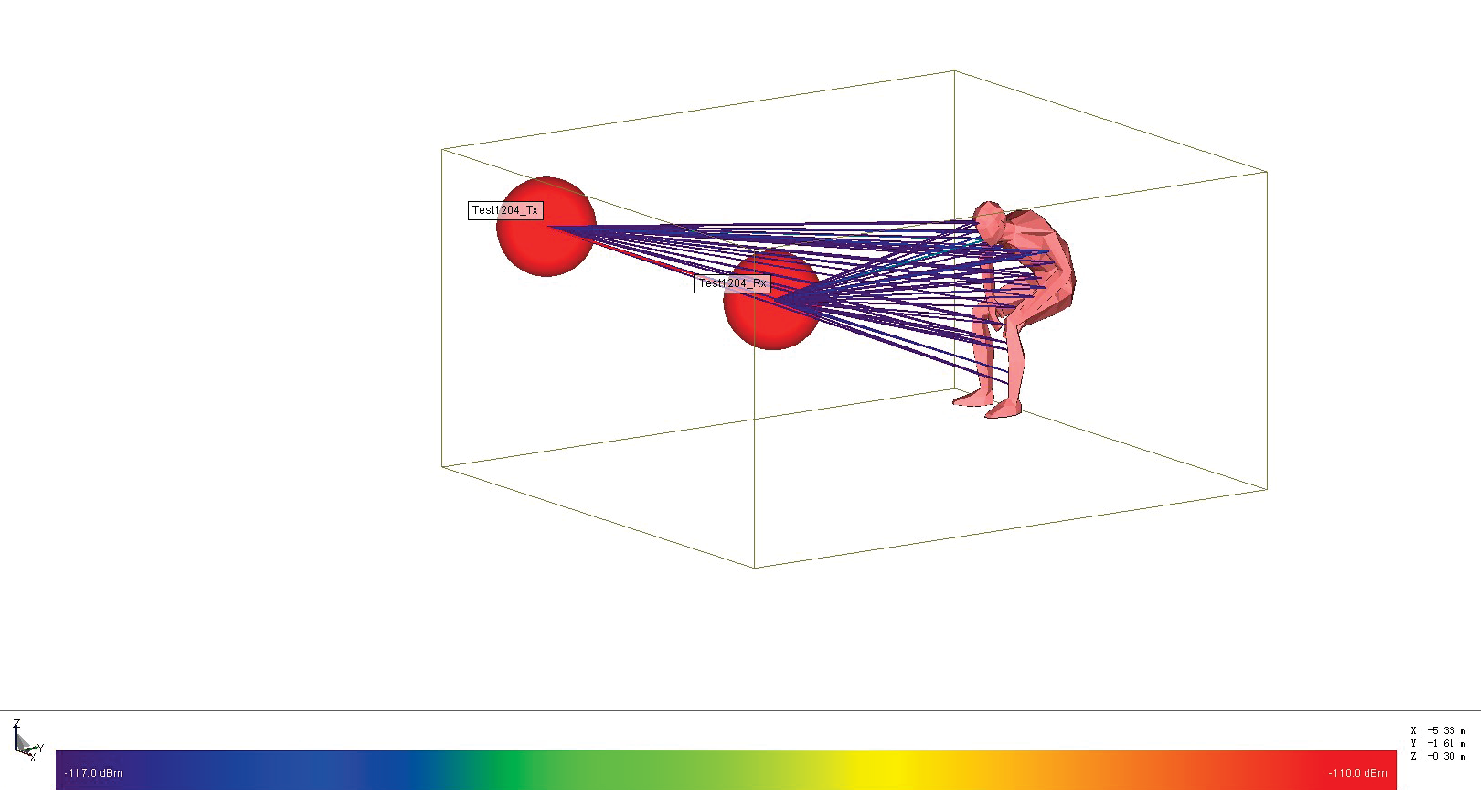}
    \caption{A slice of the ray tracing simulation of a person sitting-down.}
    \label{fig:rayact}
\end{figure}

\subsubsection{Deterministic cluster modeling}
\begin{figure}[t]
    \centering
    \subfigure[]
    {
        \includegraphics[width=0.46\linewidth]{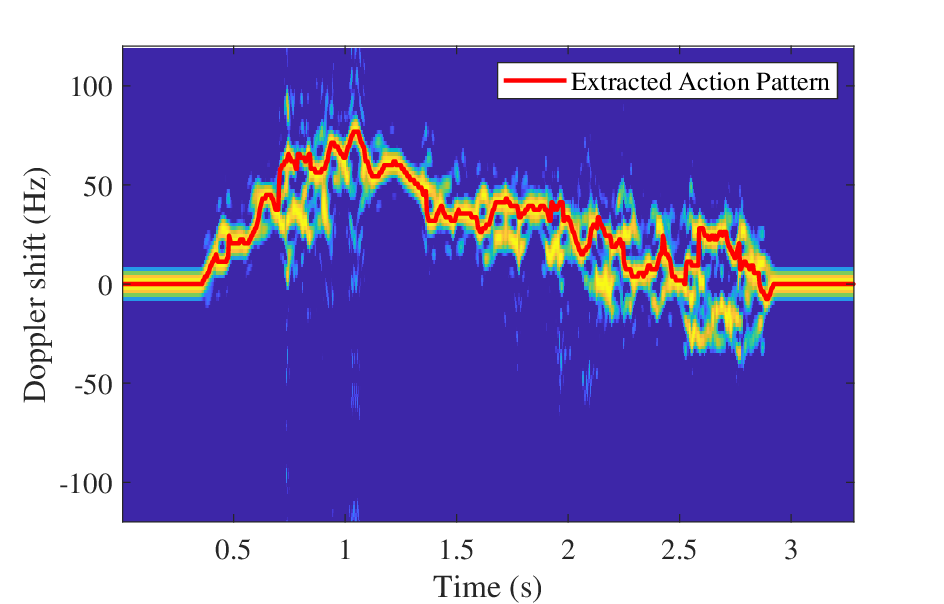}
            \label{fig:stand_ray}
    }
    \subfigure[]
    {
        \includegraphics[width=0.46\linewidth]{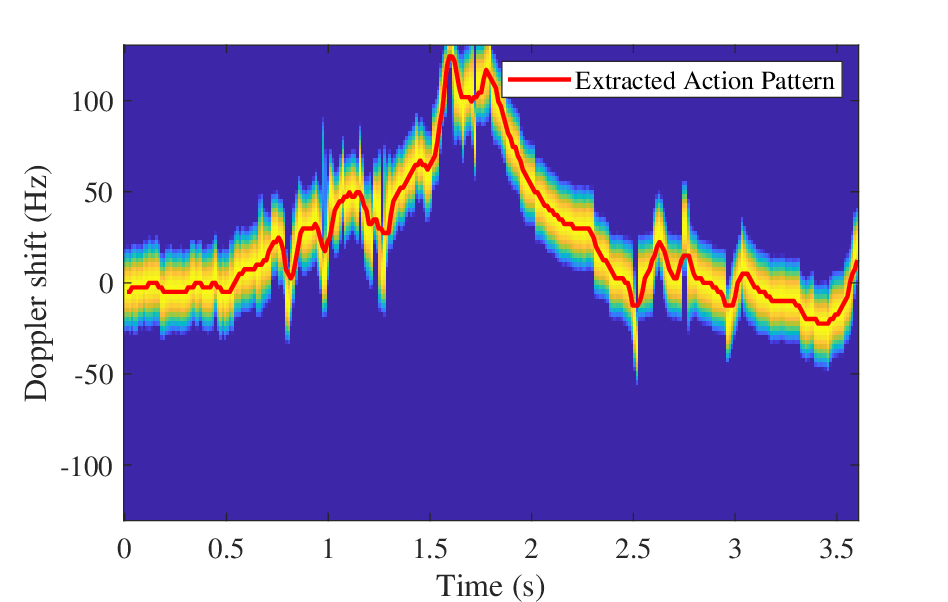}
            \label{fig:stand_exp}
    }
    \subfigure[]
    {
        \includegraphics[width=0.46\linewidth]{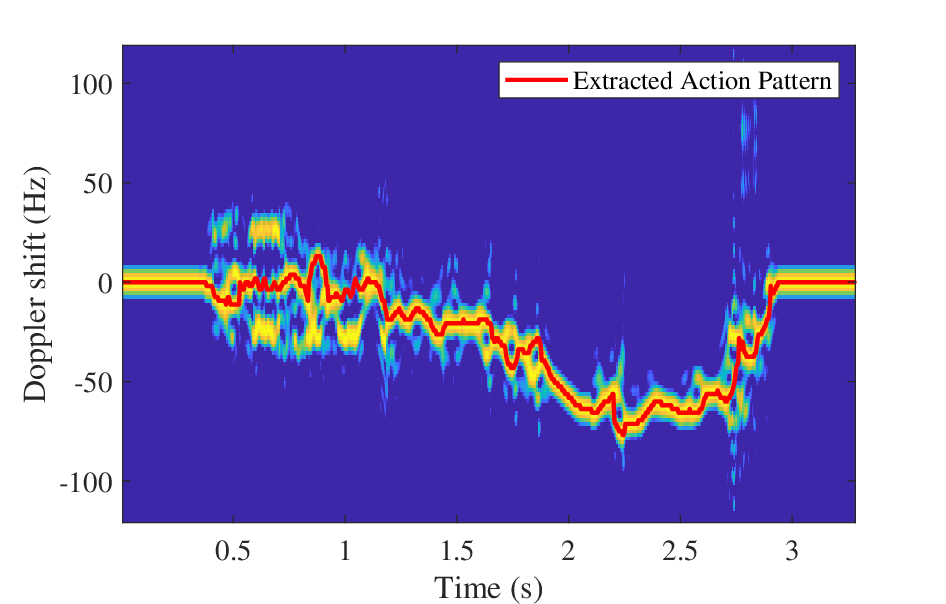}
         \label{fig:sit_ray}
    }
    \subfigure[]
    {
        \includegraphics[width=0.46\linewidth]{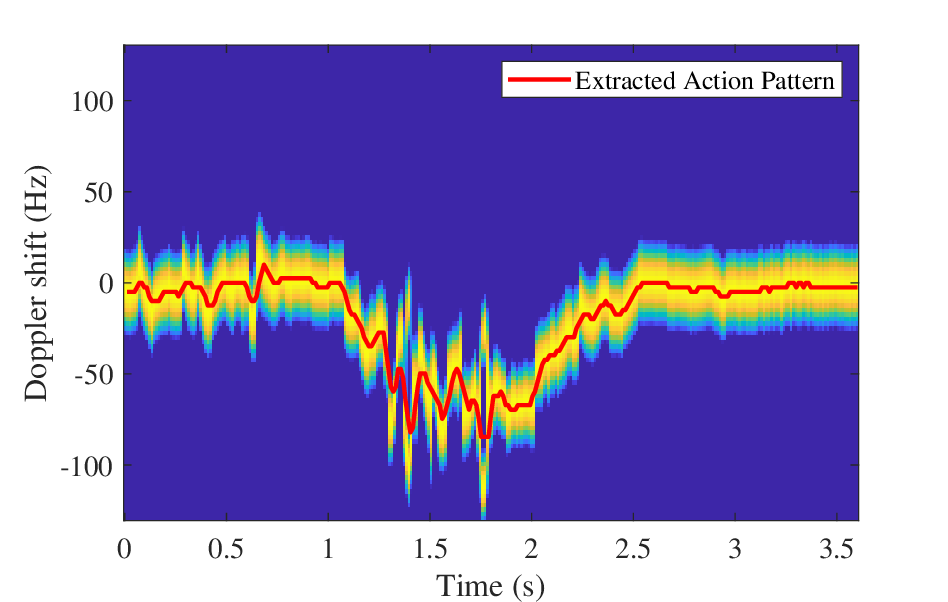}
         \label{fig:sit_exp}
    }
    \caption{The spectrogram of different actions using deterministic modeling. (a) and (b) are the patterns of standing up extracted from ray tracing and experiment respectively; (c) and (d) are the patterns of sitting down extracted from ray tracing and experiment respectively.}
    \label{fig:action}
    \vspace{-0em}
\end{figure}

In the second simulation, deterministic modeling for target cluster is studied.
More specifically, two human behaviors, namely standing-up and sitting-down, are simulated by ray tracing method. In this simulation, the transmitter and receiver are separately placed and a human model is placed in front of the transmitter and receiver. A slice of the ray tracing simulation in the case of sitting-down is shown in Fig. \ref{fig:rayact}, which can be used to show the simulation setup. To simulate the whole behavior, the process of the action is decomposed into hundreds of frames with a rate of 480 frames per second (fps). In each frame, the corresponding 3D person model is utilized for a round of ray tracing simulation as one slot. The CIR of the target cluster is constructed by collecting all the rays from the simulation results at that specific slot.
The simulation results of rays, including the power, time delay, and angle of each ray, can provide sufficient parameters to build up a deterministic target cluster model.
The ray tracing based deterministic model is inserted into our ISAC channel model for further sensing evaluation.
The sensing algorithm follows the following procedure. We first extract the CIR of target and then analyze the spectrogram of the behavior. The extracted spectrograms for standing-up and siting-down are show in Fig. \ref{fig:stand_ray} and \ref{fig:sit_ray}, respectively. The behavior patterns for both standing-up and siting-down match the spectrogram of radar measurements for these two behaviors in \cite{act}.  We can also compare the results between simulations and measurements and see consistent patterns.

\subsection{Validation by Experiments}

\subsubsection{Experiment setup}

Some experiments are carried out in an indoor office (shown in Fig. \ref{fig:experiment_setup_empty}) to verify a few key features of the proposed channel modeling framework.
The system is operating on the central frequency of 28 GHz with 500 MHz bandwidth.  Particularly, it utilizes OFDM waveform with 1824 subcarriers and a carrier interval of 270 kHz. The transmission power is set to 0 dBm. The method in \cite{li2023integrating}  is utilized for the system calibration for sensing.


\subsubsection{Power of target cluster}
\begin{figure}[t]
    \centering
    \includegraphics[width=0.7\linewidth]{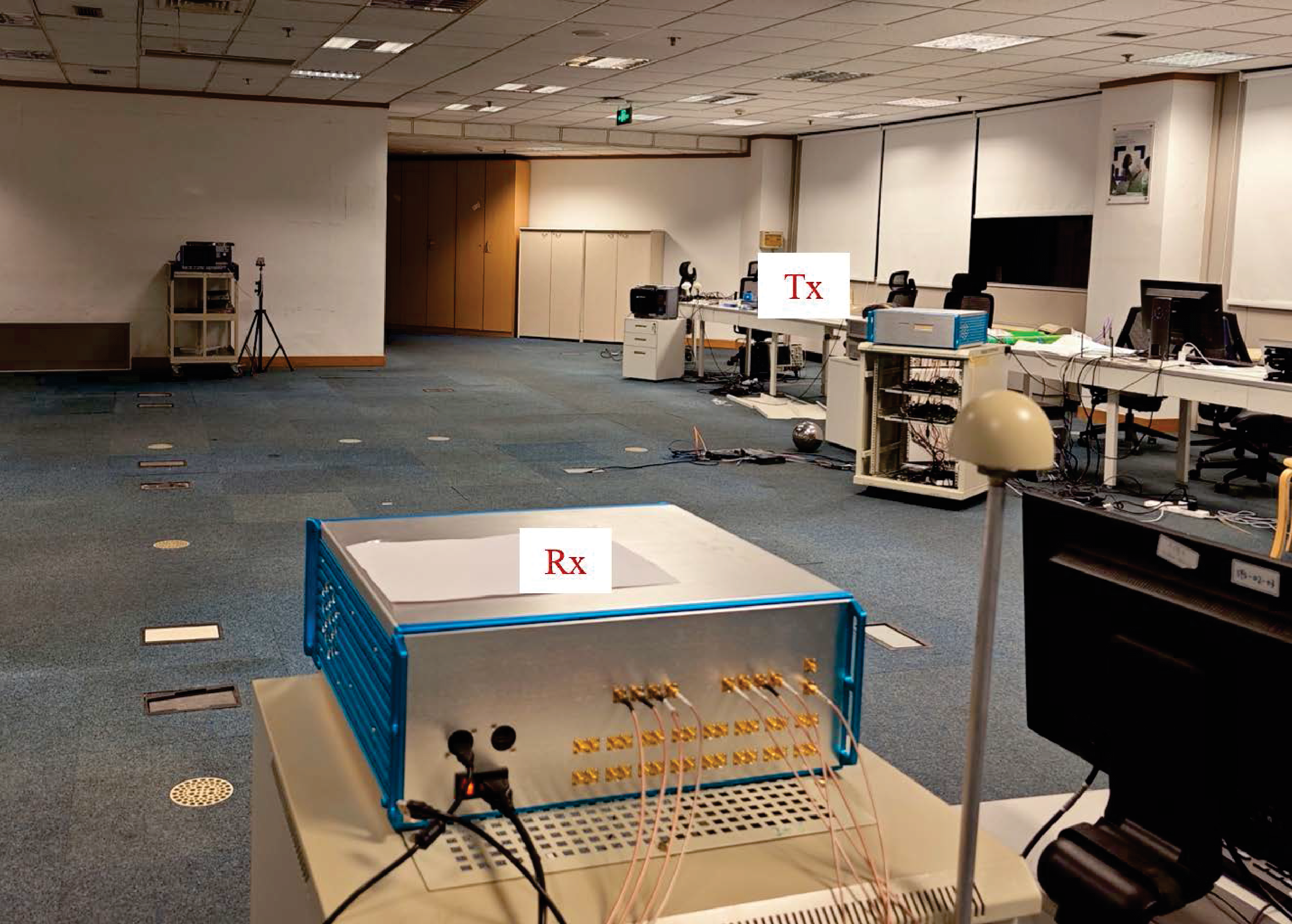}
    \caption{Experiment environment.}
    \label{fig:experiment_setup_empty}
    \vspace{-0em}
\end{figure}

\begin{figure}[t]
    \centering
    \includegraphics[width=0.9\linewidth]{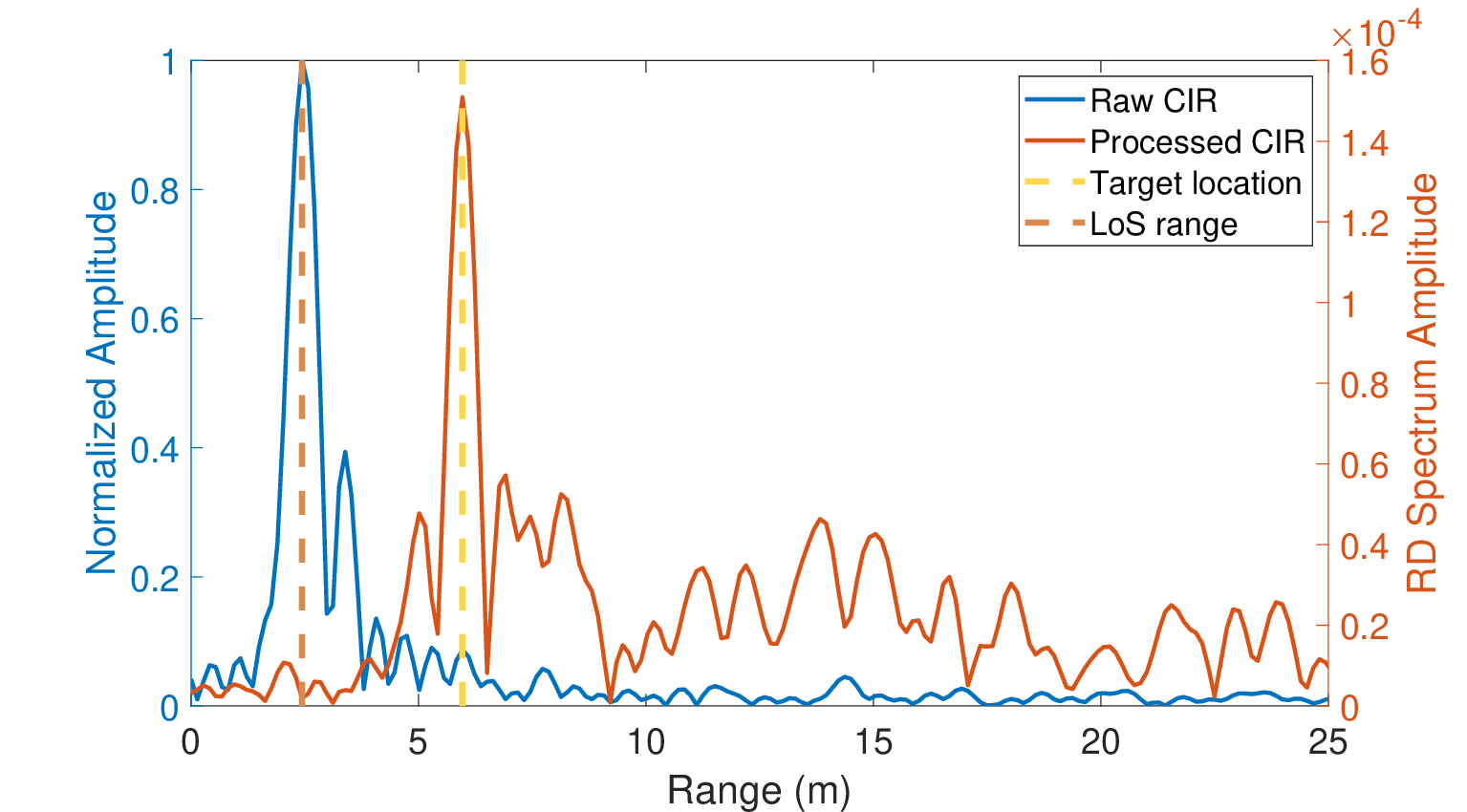}
    \vspace{-0em}
    \caption{The raw CIR and radar processed CIR for the weak target.}
    \label{fig:weakCIR}
    \vspace{-0em}
\end{figure}

In the first experiment, a person stands in the room as a sensing target as in Fig. \ref{fig:experiment_setup_empty}.
With the collected data, the raw CIR and also the processed CIR with radar sensing processing are extracted. Particularly, the raw CIR is obtained by an inverse Fourier transform of the channel frequency response, where the amplitude of CIR taps is normalized by the LoS tap. Since the silhouette of human body is irregular and the surface material of the clothes is rough, the reflection energy from the human body is very weak. The raw CIR and radar processed CIR for the weak target/human is shown in Fig. \ref{fig:weakCIR}, where the solid blue line shows the raw CIR, the dotted red line shows the accurate range of LoS path, the solid red line shows the processed CIR for target (the influence of LoS path and other strong path is removed), and the dotted yellow line shows the accurate range of target location. From the original raw CIR, we can see a strong LoS path. However, at the accurate target range, it is not straightforward to directly identify a target, since it seems a sidelobe of the strongest path. The raw CIR is extracted by the method \cite{li2023integrating}.
The raw CIR is then processed by removing the LoS path and also the reflection from the environment and conducting coherent processing using a periodogram-based algorithm \cite{braun2014ofdm} with 50 symbols to get the Range-Doppler map. In the processed CIR, which is extracted from the peak of the Range-Doppler map, the CIR tap of the target can be clearly figured out, which means that such a weak target can be successfully sensed.
Moreover, we can find out that power of the human reflection path is 30.8 dB lower than that of the LoS path, which is
weaker than the threshold set in the standard communication model in \cite{38901}.
Therefore, this experiment result also reveals the necessity of adding more clusters and setting a lower threshold for removing clusters in the bistatic ISAC channel model.



\subsubsection{Deterministic cluster modeling}
In the second experiment, deterministic modeling for target cluster is studied.
More specifically, the CIRs from experiment results of behavior recognition for two behaviors are utilized in this validation.
The CIRs from the standing-up and sitting-down behavior are first extracted from the experiments and then substituted for the target cluster in step 8d to get the final ISAC channel. With this ISAC channel, we can further validate its sensing capability for behavior recognition. The Doppler spectrograms for the two cases are shown in Fig. \ref{fig:stand_exp} and \ref{fig:sit_exp}. From the spectrogram, we can easily identify the behavior and the behaviors of standing-up and sitting-down exhibit opposite spectral patterns, which is consistent with the spectrogram of radar measurements in \cite{act}. This indicates that replacing the target cluster in step 8d with the deterministic target model is able to characterize the features of complicated sensing task.



\subsection{Compatibility with 3GPP Communication Channel Model and Evaluation for Sensing}
\subsubsection{Simulation setup} To verify the compatibility of our proposed ISAC channel with communication evaluation, we compare the communication bit-error-rate (BER) performance between our channel model and the standard communication channel model. The generation of the standard communication channel follows the procedure in 3GPP TR 38.901, while our bistatic ISAC channel generation follows our proposed procedure.
In the standard communication channel, the number of generated clusters is $N=12$, while in the ISAC channel, $N_{\text{ISAC}}=24$. In addition, the cluster removing power threshold for the standard communication channel is set to be $-25$ dB, while the threshold in the ISAC channel is lowered to $-50$ dB.
Moreover, three clusters are randomly selected as targets in each run, and statistical modeling is applied in the bistatic ISAC channel. Their velocities are set to follow the normal distribution with the standard deviation and the mean of $1$ m/s, $-10$ m/s, and $30$ m/s respectively to feature the targets of a human, a bike, and a car.
Other system parameters for simulations are listed in Table \ref{tab:para}. In addition, the block-type RS is deployed every 7 OFDM symbols, and the least square (LS) channel estimator with the spline interpolation is used for both cases.

\begin{table}[t]
    \centering
    \renewcommand{\arraystretch}{1.2}
    \renewcommand{\tabcolsep}{1.2mm}
    \caption{Simulation Parameters}
    \label{tab:para}
    \begin{tabular}{cc}
    Parameter & Value\\
    \hline
    Central frequency $f_c$ & 28 GHz \\
    Number of subcarriers & 792 \\
    Subcarrier interval $\Delta f$ & 120 kHz \\
    Symbol duration & $8.92 \times 10^{-6}$ ms \\
    Communication scenario & UMi - street canyon \\
    Sensing scenario &  target localization \\
    \hline
    \end{tabular}
    \vspace{-0em}
\end{table}

\begin{figure}[t]
    \centering
    \includegraphics[width=0.8\linewidth]{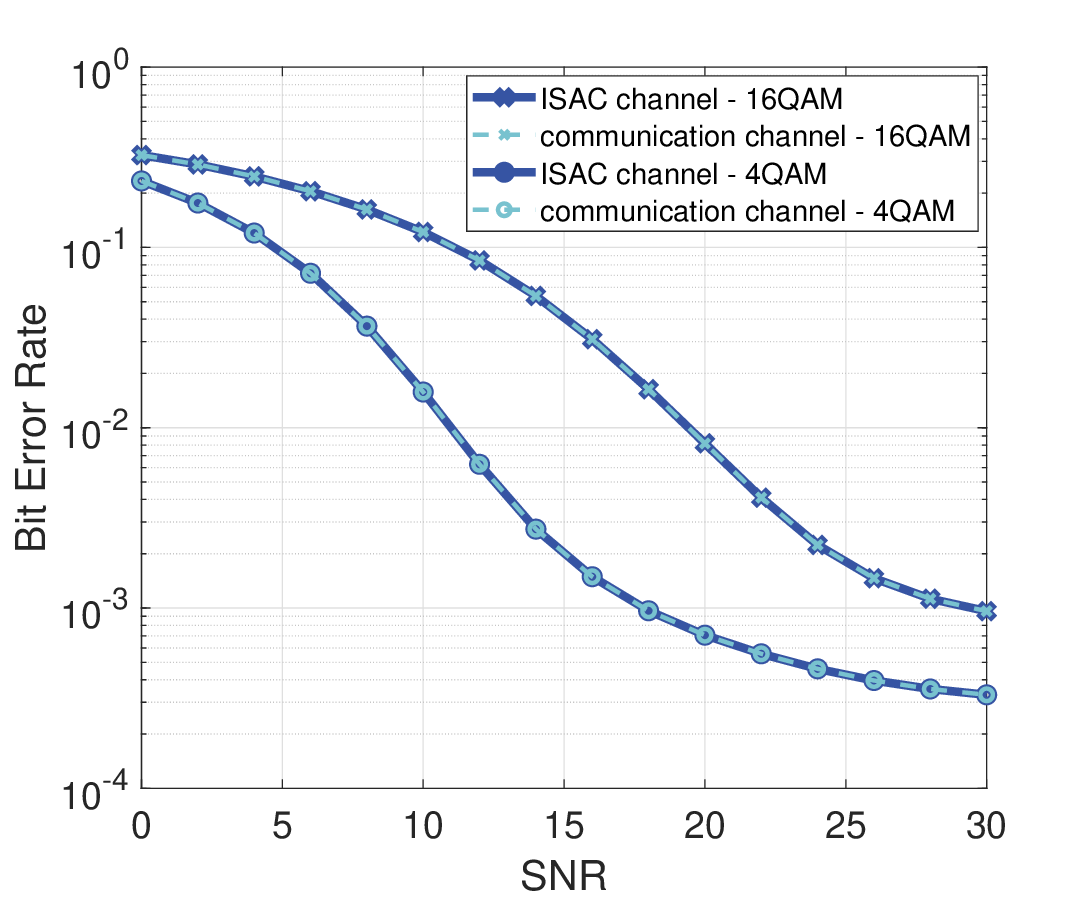}
    \caption{Comparison of the BER between the ISAC channel and 3GPP communication channel.}
    \label{fig:ber}
    \vspace{-0em}
\end{figure}

\subsubsection{Communication performance}
The BER results of the proposed ISAC channel and the 3GPP communication channel for different modulations are illustrated in Fig. \ref{fig:ber}. The result depicts that the BER performance of the ISAC channel aligns closely with that of the 3GPP communication channel, which indicates that the ISAC channel is capable of evaluating the communication function as the original communication channel, and the proposed bistatic ISAC channel is compatible with the current communication channel for communication performance evaluation. In other words, it is feasible to use the proposed unified bistatic ISAC channel model to evaluate both the communication and sensing performance.

\begin{figure}[t]
    \centering
    \subfigure[detection probability]{
	\includegraphics[width=0.8\linewidth]{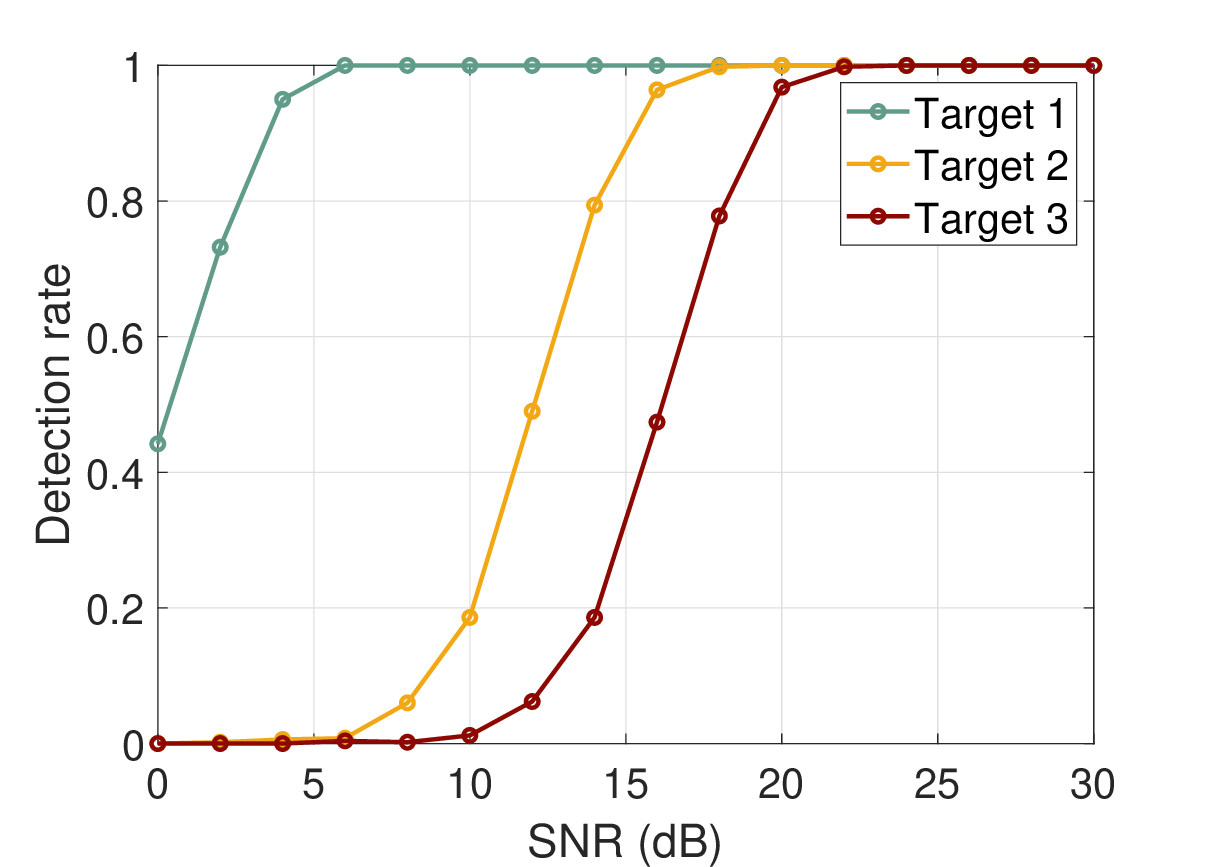}
    	\label{fig:detection}
    }
    \subfigure[range estimation error]{
        \includegraphics[width=0.8\linewidth]{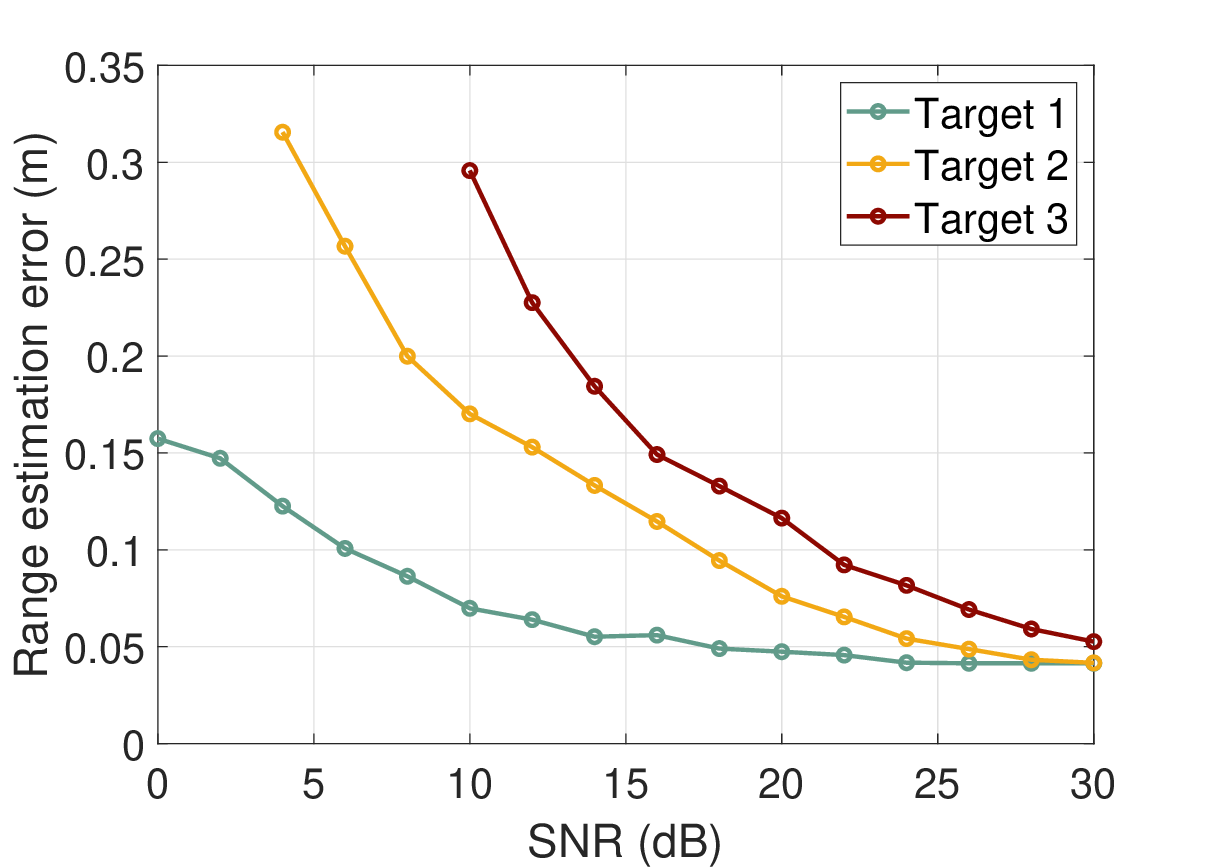}
        \label{fig:range}
    }
    \caption{Sensing evaluation study under statistic modeling: (a) detection probability, (b) range estimation error.}
    \label{fig:sensing}
    \vspace{-0em}
\end{figure}

\subsubsection{Sensing evaluation via statistical cluster modeling}
In this simulation, a case study of sensing performance evaluation under the proposed bistatic ISAC channel model framework is provided. More specifically, we use the above described simulation setup in this subsection to generate an ISAC channel. In this case study, the following detailed setup is used for target clusters. Three target clusters are considered. Point target model is used for each cluster and only the type 1 reflection model is adopted, so that the range of each target can be directly estimated. Three targets are randomly selected from the generated cluster and the power of the three targets is  -27.76 dB (target 1), -39.41 dB (target 2), and -43.41 dB (target 3) lower than the LoS path.

The periodogram-based sensing algorithm \cite{braun2014ofdm} is performed for target detection and range estimation, where constant false alarm rate (CFAR) detector with false alarm rate $10^{-5}$ is adopted. The radar processing gain is about 30 dB. We change the total power of the ISAC channel, i.e., the channel SNR, and evaluate the sensing performance of detection rate and range estimation error. For a fixed SNR, we run 1000 times with random noise generation to collect the results. The simulation results are shown in Fig. \ref{fig:sensing}. As we can see from Fig. \ref{fig:detection}, the strongest target (namely target 1) can be reliably detected when the channel SNR is larger than 6 dB. If the channel SNR is larger than 22 dB, all the three targets can be reliably detected. This example also shows the necessity to give a lower removing threshold in bistatic ISAC modeling. The range of the three targets also can be estimated with high precision as shown in Fig. \ref{fig:range}.


\section{Conclusion}

In this paper, a bistatic ISAC channel modeling framework was proposed under the 3GPP TR 38.901 standard. The proposed framework extended the current framework with wireless sensing capability. Main modifications in the proposed framework included setting both communication and sensing scenarios, generating more clusters, retaining more clusters with weaker power, and modeling the target clusters in a deterministic or statistical manner according to the sensing scenario, generating channel coefficients for sensing cluster. Ray tracing simulation and experiment implementation were carried out to validate the proposed modeling framework. Simulation results also showed the compatibility with current channel model of using the ISAC channel for communications and demonstrated the capability for sensing performance evaluation.

However, this paper only proposes the extended channel model framework for bistatic ISAC. The determination of detailed parameters (such as the exact number of generated clusters and removing threshold for each ISAC scenario, the selection of sensing clusters, the target modeling and reflection modeling for each sensing application) requires more comprehensive experiment measurements for each scenario, which is subject to future studies.

\bibliographystyle{IEEEtran}
\bibliography{reference}

\begin{IEEEbiography}[{\includegraphics[width=1in,height=1.25in,clip,keepaspectratio]{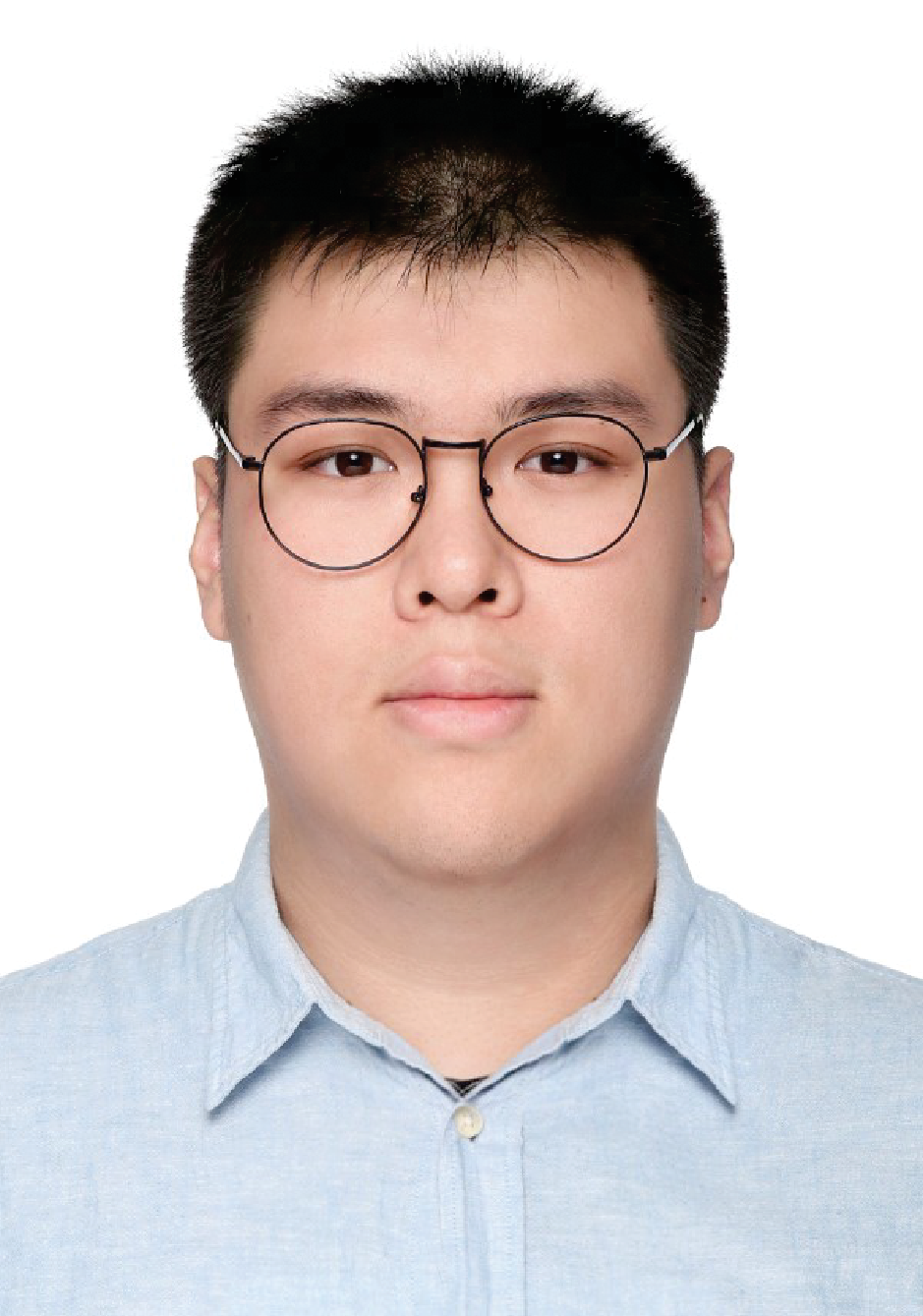}}]{Chenhao Luo}
received the B.S. degree in electrical and computer engineering from Shanghai Jiao Tong University, Shanghai, China, in 2022. He is currently pursuing the master’s degree in information and communication engineering with the University of Michigan–Shanghai Jiao Tong University Joint Institute, Shanghai Jiao Tong University. His research interests include integrated sensing and communications, channel modeling, and mmWave imaging.
\end{IEEEbiography}

\begin{IEEEbiography}[{\includegraphics[width=1in,height=1.25in,clip,keepaspectratio]{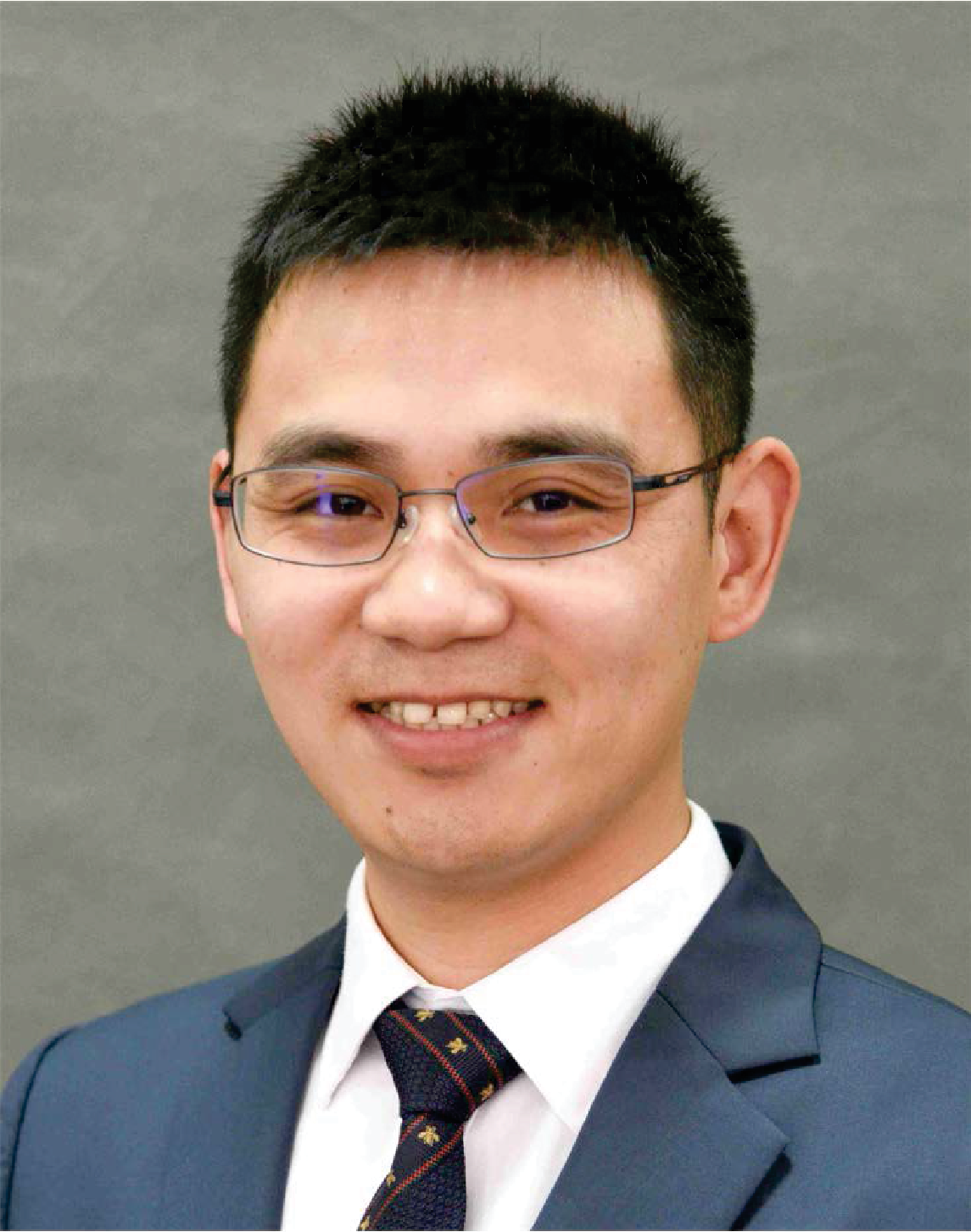}}]{Aimin Tang}
(Member, IEEE) received the B.S. and Ph.D. degrees in information and communication engineering from Shanghai Jiao Tong University, Shanghai, China, in 2013 and 2018, respectively.
He was a joint Ph.D. student at the University of Washington in 2016. He received the Best Paper Award of IEEE International Symposium on Personal, Indoor and Mobile Radio Communications (PIMRC) 2021. He is currently a Research Assistant Professor with the University of Michigan-Shanghai Jiao Tong University (UM-SJTU) Joint Institute, Shanghai Jiao Tong University. His current research interests include B5G/6G networks, integrated sensing and communications, and full-duplex communications. 
\end{IEEEbiography}

\begin{IEEEbiography}[{\includegraphics[width=1in,height=1.25in,clip,keepaspectratio]{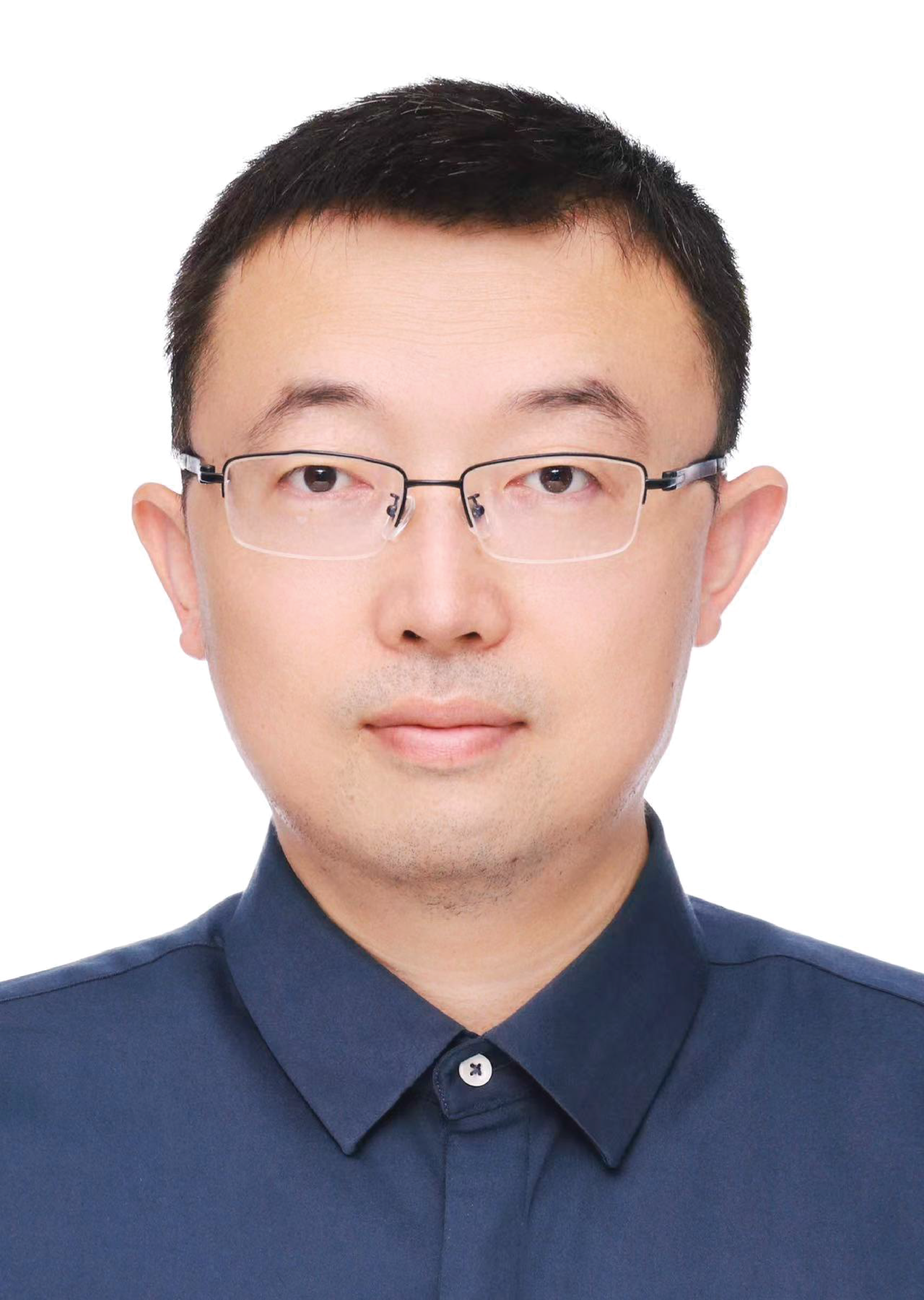}}]{Fei Gao}
(Senior Member, IEEE) received the B.S. and Ph.D. degrees in electronic engineering from Xidian University, Xi'an, China, in 2010 and 2015, respectively.
He was a joint Ph.D. student at the University of California San Diego from 2012 to 2014. Currently, he is the director of perception and automation R$\&$D department of Nokia Bell Labs China. Dr. Gao's research interests include smart antennas, intelligent reflective surfaces, indoor positioning, and human perception. The results of his research on high-precision localization using WiFi and BLE technologies have been extensively integrated into various industry products. He has published more than 60 research papers and is currently an industry editor of ISAC-Focus.
\end{IEEEbiography}

\begin{IEEEbiography}[{\includegraphics[width=1in,height=1.25in,clip,keepaspectratio]{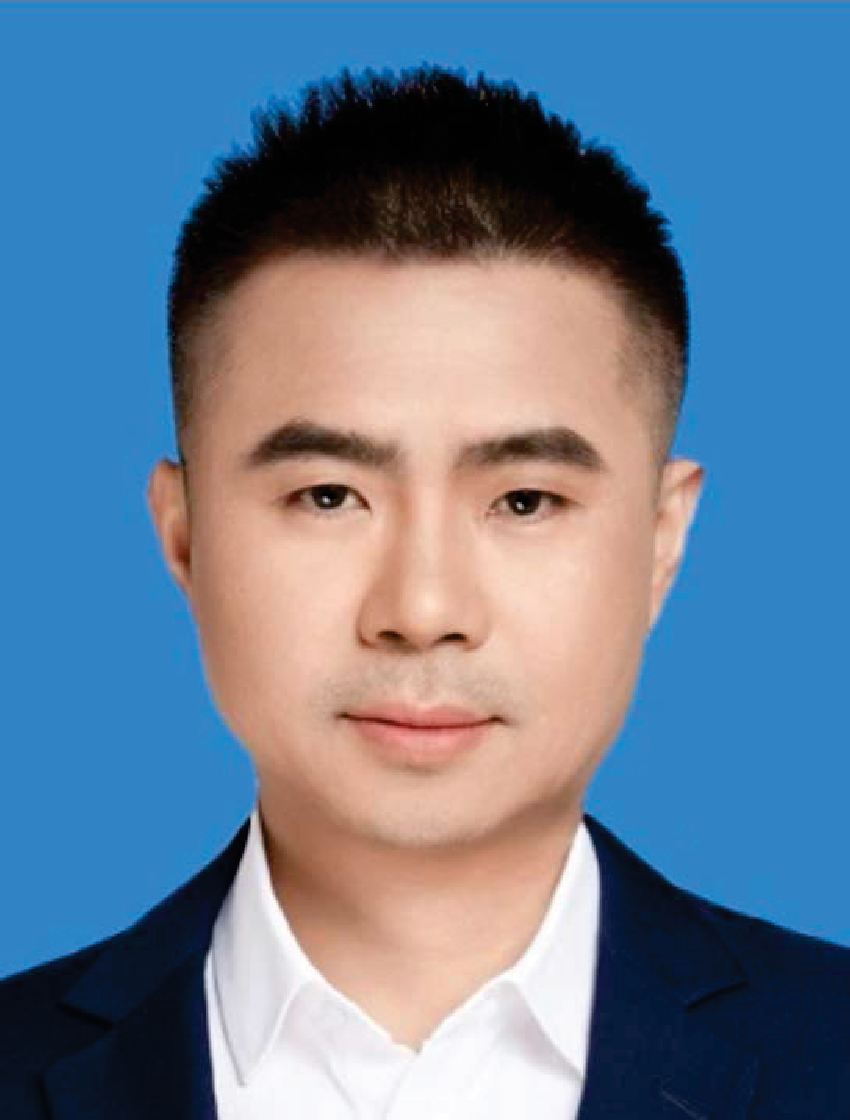}}]{Jianguo Liu}
(Member, IEEE) received the Ph.D. degree in School of Information Science and Engineering from Southeast University, Nanjing, China in 2008. He is currently a Research Scientist in Nokia Bell Labs China. His current research interests include integrated sensing and communication, NR positioning, and Ambient IoT Communications.
\end{IEEEbiography}

\begin{IEEEbiography}[{\includegraphics[width=1in,height=1.25in,keepaspectratio]{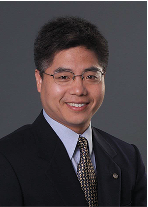}}]
{Xudong Wang} (Fellow, IEEE) received the Ph.D. degree in electrical and computer engineering from the Georgia Institute of Technology in 2003. He is currently the John Wu and Jane Sun Chair Professor of engineering at the UM-SJTU Joint Institute, Shanghai Jiao Tong University. He is also an affiliate Professor with the Department of Electrical and Computer Engineering, University of Washington. His research interests include wireless communication networks, distributed machine learning, and joint communications and sensing. He was a voting member of IEEE 802.11 and 802.15 Standard Committees. He served as an Associate Editor for IEEE Transactions on Mobile Computing, IEEE Transactions on Vehicular Technology, Ad Hoc Networks (Elsevier), and {China Communications}. He was also a Guest Editor of several international journals and general chair or TPC co-chair for several international conferences.
\end{IEEEbiography}

\end{document}